\documentclass[preprint,showpacs,amsmath,amssymb,aps,prd,nofootinbib]{revtex4}
\usepackage{epsfig}
\usepackage{axodraw}
\usepackage{color}
\begin{document}
\draft
\title{\mbox{}\\[10pt]
 Non-zero $U_{e3}$, TeV-Leptogenesis through $A_{4}$ symmetry breaking}

\author{Y.~H.~Ahn$^1$\footnote{E-mail:
        yhahn@phys.sinica.edu.tw},Chian-Shu~Chen$^{1,2}$\footnote{E-mail: chianshu@phys.sinica.
       edu.tw}}

\affiliation{1.~Institute of Physics, Academia Sinica, Taipei, Taiwan 115, ROC.}
\affiliation{2.~National Center for Theoretical Sciences (South), Tainan, Taiwan 701, ROC.}




\begin{abstract}
We consider an effective theory with an $SU(2)\times U(1)\times A_{4}\times Z_{2}\times Z_{4}$ symmetry and investigate the possibility of a linking  TeV-leptogenesis with a reactor angle $|U_{e3}|$ through $A_{4}$ symmetry breaking which is at a scale higher than electroweak scale under the framework of radiative seesaw. It has been shown that tri-bimaximal(TBM) can be obtained by forging vacuum expectation value (VEV) alignment of the $A_{4}$. Especially, one $A_{4}$ triplet scalar field with cut-off scale $\Lambda$ is added in neutrino Yukawa sector, which is responsible for the deviation of the exact TBM, to explain leptogenesis as well as a non-zero $|U_{e3}|$. Above the scale of $\Lambda$ the leptonic Yukawa sectors will lead to the exact TBM. We analyze possible spectrums of light neutrinos and their flavor mixing angles corresponding to heavy Majorana neutrino mass ordering, and show that non-resonance leptogenesis at TeV-scale constrained by low energy data is achievable, both analytically as well as numerically. We show that only normal hierarchical spectrum of light neutrino would be strongly favored by the current Wilkinson Microwave Anisotropy Probe (WMAP) data, and also show that a relatively large $|U_{e3}|$ corresponds to the value of baryon asymmetry $6.2\times10^{-10}$.

\noindent
\end{abstract}
 \maketitle
\section{Introduction}
Recent analysis on the knowledge of neutrino oscillation parameters, which makes desirable a neutrino texture going beyond the mere fitting procedure, has shown in Table-\ref{tab:data},
\begin{table}[th]
\begin{center}
\begin{tabular}{|c|c|c|c|c|c|} \hline
 & $\Delta m^{2}_{\rm Sol}/10^{-5}\mathrm{\ eV}^2$ & $\sin^2\theta_{12}$ & $|U_{e3}|$ & $\sin^2\theta_{23}$ &
$\Delta m^{2}_{\rm Atm}/10^{-3}\mathrm{\ eV}^2$ \\ \hline \hline
Best-fit        &     7.67     &  0.312          &  0.126          &
0.466          &  2.39 \\ \hline
$1\sigma$ & $7.48 - 7.83$ & $0.294 - 0.331$  & $0.077 -
0.161$  & $0.408 - 0.539$  & $2.31 - 2.50$ \\ \hline
$3\sigma$ & $7.14 - 8.19$ & $0.263 - 0.375$  & $<0.214$
& $0.331 - 0.644$  & $2.06 - 2.81$ \\ \hline
\end{tabular}
\caption{\label{tab:data}Current best-fit values as well as 1 and $3\sigma$ ranges of the oscillation parameters \cite{bari}.}
\end{center}
\end{table}
which at $3\sigma$ fully compatible with the TBM pattern
 \begin{eqnarray}
  \sin\theta_{12}=\frac{1}{\sqrt{3}},~~~~~~\sin\theta_{23}=\frac{-1}{\sqrt{2}},~~~~~~\sin\theta_{13}=0~.
  \label{tibi1}
 \end{eqnarray}
However, the recent analysis based on global fits of the available data gives us hints for $\theta_{13}>0$ at $1\sigma$\cite{Fogli:2009ce,nudata}. Although neutrinos have gradually revealed their properties in various experiments since the historic Super-Kamiokande confirmation of neutrino oscillations \cite{Fukuda:1998mi}, properties related to the leptonic CP violation are completely unknown yet. In addition, the large mixing values of $\theta_{\rm sol}\equiv\theta_{12}$ and $\theta_{\rm atm}\equiv\theta_{23}$ may be telling us about some new symmetries of leptons that are not present in the quark sector and may provide a clue of the nature among quark-lepton physics beyond the standard model.
The most popular discrete symmetry $\mu-\tau$ symmetry have some success in describing the mass and mixing pattern in the leptonic sector~\cite{mutau}. Nevertheless, E.Ma and G.Rajasekaran~\cite{Ma:2001dn} have introduced for the first time the $A_{4}$ symmetry to avoid mass degeneracy of $\mu$ and $\tau$ under $\mu-\tau$ symmetry. In a well-motivated extension of the standard model through the inclusion of $A_{4}$ discrete symmetry the TBM pattern comes out in a natural way in the work of \cite{He:2006dk}.
Models of $A_4$ symmetry implemented with grand unification~\cite{Altarelli:2008bg}, supersymmetry~\cite{Bazzocchi:2007na}, and extra dimensions\cite{Altarelli:2005yp,Altarelli:2006kg} are also investigated extensively in literatures.

On the other hand, the observed baryon asymmetry in our universe (BAU) can be explained by the mechanism of leptogenesis\cite{Fukugita:1986hr, Langacker:1986rj}. $A_{4}$ models realized on type-I seesaw lead to vanishing leptonic CP-asymmetries responsible for leptogenesis due to the combination of Dirac neutrino Yukawa coupling matrix  $Y^{\dag}_{\nu}Y_{\nu}$ being proportional to the unit matrix. A common proposal to address the possibility of leptogenesis in $A_{4}$ models is adding $A_4$ soft breaking terms into Lagrangian such that the deviation of TBM as well as CP-asymmetries responsible for leptogenesis can be generated \cite{Adhikary:2008au}, while in Ref.\cite{Branco:2009by, Jenkins:2008rb, Lin:2009bw} the authors considered higher dimensional operators based on an effective theory. Instead of that, we add one 5-dimensional effective operator with respect to $\Lambda$ under $SU(2)\times U(1)\times A_4\times Z_2\times Z_4$, and $A_{4}\times Z_{4}$ symmetries are broken after the assuming scalars develop VEVs with {\it ad hoc} constraints in the potential\footnote{see, more details in Appendix}, which opens the possibility to study an attractive mechanism of leptogenesis and to connect this with low-energy observables without contradicting $1\sigma$ results \cite{bari}.

Besides the mystery of the mixing pattern, tiny neutrino mass is one of the most challenging problem beyond Standard Model. Recently, E.Ma introduced the so-called radiative seesaw mechanism \cite{Ma:2006km} where the neutrino masses are generated through one-loop mediated by a new Higgs doublet and right-handed neutrinos obeying an additional $Z_2$ symmetry. In this paper, we address the possibility of an linking between TeV-leptogenesis and non-zero $\theta_{13}$ through symmetry breaking of $A_{4}$ in a radiative seesaw mechanism. Our starting point is an effective Lagrangian with an $A_{4}\times Z_2\times Z_{4}$ symmetry which is broken by the VEV of $SU(2)_{L}\times U(1)_{Y}$ singlet scalar fields at a scale higher than the electroweak scale. In addition to this, we assign a $Z_{2}$-odd quantum number to a leptonic Higgs doublet $\eta=(\eta^{+},\eta^{0})$ and three right-handed singlet fermions $N_{i}$ while all the standard model particles are $Z_{2}$-even. After electroweak symmetry breaking, the $Z_{2}$ symmetry is exactly conserved and $\eta$ will not develop a VEV, that is $\langle\eta^{0}\rangle=0$, while the standard Higgs boson get a VEV, which means the Yukawa coupling corresponding to $Z_{2}$-odd Higgs doublet will not generate the Dirac mass terms in neutrino sector. Thus, the usual seesaw mechanism does not work any more and we naturally have a good candidate of dark matter (DM) corresponding to the lightest $Z_{2}$-odd particle or Large Hadron Collider (LHC) signals through the standard gauge interactions in our model. The assigned leptonic flavor symmetry will lead us to the TBM, and for both its deviation and leptogenesis to be explained one $A_4$ triplet scalar field with cutoff scale $\Lambda$ is introduced in leptonic sector. We analyze possible spectrums of light neutrinos and their flavor mixing angles. And we show that non-resonance leptogenesis at TeV-scale constrained by low energy data is achievable.

The paper is organized as follows. In the next section, we present the particle content together with the flavor symmetry of our model. In Sec.~$\textrm{III}$, we show the neutrino masses are generated in 1-loop level and how the parameters are constrained by the low energy neutrino oscillation data. Sec.~$\textrm{IV}$ analyzes leptogenesis included flavor effects in each heavy right-handed neutrino spectrum. Then we give the conclusion in Sec.~$\textrm{V}$, and in Appendix we briefly discuss the vacuum alignment challenge.

\section{flavor $A_{4}$ symmetry and a discrete symmetry $Z_{2}\times Z_{4}$}

Unless flavor symmetries are assumed, particle masses and mixings are generally undetermined in gauge theory. To understand the present neutrino oscillation data we consider $A_{4}$ flavor symmetry for leptons, and simultaneously the existence of LHC signal and the baryon asymmetry of the Universe to be explained at TeV scale we also introduce an extra discrete symmetry $Z_{2}$ in a radiative seesaw~\cite{Ma:2006fn}.
Especially, we introduce a 5-dimensional operator in the lagrangian which is invariant under $A_4\times Z_2\times Z_{4}$ to have non-zero low energy CP violation in neutrino oscillation and non-zero high energy cosmological CP violation which is responsible for BAU.
Here we recall that $A_{4}$ is the symmetry group of the tetrahedron and the finite groups of the even permutation of four objects.
Its irreducible representations contain one triplet ${\bf 3}$ and three singlets ${\bf 1}, {\bf 1}', {\bf 1}''$ with the multiplication rules are ${\bf 3}\otimes{\bf 3}={\bf 3}_{s}\oplus{\bf 3}_{a}\oplus{\bf 1}\oplus{\bf 1}'\oplus{\bf 1}''$, and ${\bf 1}'\otimes{\bf 1}'={\bf 1}''$. Let's denote two $A_4$ triplets, $(a_{1}, a_{2}, a_{3})$ and $(b_{1}, b_{2}, b_{3})$, then we have
 \begin{eqnarray}
  {\bf 3}_{\rm s} &\sim& (a_{2}b_{3}+a_{3}b_{2}, a_{3}b_{1}+a_{1}b_{3}, a_{1}b_{2}+a_{2}b_{1})~,\nonumber\\
  {\bf 3}_{\rm a} &\sim& (a_{2}b_{3}-a_{3}b_{2}, a_{3}b_{1}-a_{1}b_{3}, a_{1}b_{2}-a_{2}b_{1})~,\nonumber\\
  {\bf 1} &\sim& a_{1}b_{1}+a_{2}b_{2}+a_{3}b_{3}~,\nonumber\\
  {\bf 1}' &\sim& a_{1}b_{1}+\omega a_{2}b_{2}+\omega^{2}a_{3}b_{3}~,\nonumber\\
  {\bf 1}'' &\sim& a_{1}b_{1}+\omega^{2} a_{2}b_{2}+\omega a_{3}b_{3}~,
 \end{eqnarray}
where $\omega=e^{i2\pi/3}$ is a complex cubic-root of unity.

The field content under $SU(2)\times U(1)\times A_{4}\times Z_{2}\times Z_{4}$ of the model is assigned as Table-\ref{reps}
\begin{widetext}
\begin{center}
\begin{table}[h]
\caption{\label{reps} Representations of the fields under $A_4 \times Z_{2} \times Z_4$ and $SU(2)_L \times U(1)_Y$.}
\begin{ruledtabular}
\begin{tabular}{ccccccccc}
Field &$\ell_{L}$&$l_R,l'_R,l''_R$&$N_R$&$\vartheta$&$\eta$&$\psi$&$\chi$&$\Phi$\\
\hline
$A_4$&$\mathbf{3}$&$\mathbf{1}$, $\mathbf{1^\prime}$,$\mathbf{1^{\prime\prime}}$&$\mathbf{3}$&$\mathbf{1}$&$\mathbf{1}$&$\mathbf{3}$&$\mathbf{3}$&$\mathbf{3}$\\
$Z_2$&$+$&$+$&$-$&$+$&$-$&$+$&$+$&$+$\\
$Z_4$&$1$&$i$&$-i$&$-1$&$i$&$1$&$-1$&$i$\\
$SU(2)_L\times U(1)_Y$&$(2,-1)$&$(1,-2)$&$(1,0)$&$(1,0)$&$(2,-1)$&$(1,0)$&$(1,0)$&$(2,-1)$\\
\end{tabular}
\end{ruledtabular}
\end{table}
\end{center}
\end{widetext}
Hence its Yukawa interactions in the lepton sector, which is invariant under  $SU(2)\times U(1)\times A_{4}\times Z_{2}\times Z_{4}$, can be written as
 \begin{eqnarray}
 {\cal L}_{\rm Yuk} &=& g_{\nu}(\bar{\ell}_{L}N_{R})_{{\bf 1}}\eta+\frac{f_{\nu}}{\Lambda}(\bar{\ell}_{L}N_{R}\psi)_{{\bf 1}}\eta+y_{R}\vartheta[\bar{N}_{R}(N_{R})^{c}]_{{\bf 1}}+\lambda_{\chi}[\bar{N}_{R}(N_{R})^{c}]_{{\bf 3}_{s}}\cdot \chi\nonumber\\
 &&+ y_{e}(\bar{\ell}_{L}\tilde{\Phi})_{{\bf 1}}l_{R}+y_{\mu}(\bar{\ell}_{L}\tilde{\Phi})_{{\bf 1}'}l''_{R}+y_{\tau}(\bar{\ell}_{L}\tilde{\Phi})_{{\bf 1}''}l'_{R}+h.c.
 \label{lagrangian}
 \end{eqnarray}
where $\tilde{\Phi}\equiv i\tau_{2}\Phi^{\ast}$ and $\Lambda$ is a cutoff scale. Note here that we add the extra symmetry $Z_{4}$ in order to prevent direct couplings of the right-handed neutrinos to $\psi$ and $\chi$, i.e $\bar{N}_{R}(N_{R})^{c}\psi$ and $\frac{f_{\nu}}{\Lambda}\bar{\ell}_{L}N_{R}\chi\eta$. We assume that above a cutoff scale $\Lambda$ there is no CP-violation term in neutrino Yukawa interaction, which for scales below $\Lambda$ is expressed in terms of 5-dimensional operator. The breaking scale of $A_{4}\times Z_{4}$ is assumed to be lower than the cutoff $\Lambda$.
In above lagrangian, each charged lepton sector has three independent Yukawa terms, all involving the $A_{4}$ triplet Higgs field $\Phi$, while the Majorana masses of right-handed neutrinos are given by two electroweak singlet $\chi$ and $\vartheta$ scalars with ${\bf 3}$ and ${\bf 1}$ representations under $A_4$. By imposing a $Z_{2}$ symmetry as showed in Table-\ref{reps}, the Yukawa terms $\bar{\ell}_{L}N_{R}\Phi+h.c$ are forbidden, and the neutral component of scalar doublet $\eta$ will not generate a VEV, $<\eta^{0}>\equiv\upsilon_{\eta}=0$. Therefore, the scalar field $\eta$ can only couple to the standard gauge bosons as well as the Dirac neutrino mass terms are vanished which means the usual seesaw does not operate anymore. However, the light Majorana neutrino mass matrix can be generated radiatively through one-loop with the help of the Yukawa interaction $\bar{\ell}_{L}N_{R}\eta$ (in which Yukawa coupling matrix is proportional to the $3\times3$ identity matrix) and $\bar{\ell}_{L}N_{R}\psi\eta$ in Eq.~(\ref{lagrangian}), we will discuss this more detail in Sec. \textrm{III}.

We assume the VEVs of $A_{4}$ triplets can be equally aligned, that is, $\langle\Phi^{0}\rangle=(\upsilon,\upsilon,\upsilon)$, the charged-lepton mass matrix can be explicitly expressed as
 \begin{eqnarray}
 m_{f}= U_{\omega}{\left(\begin{array}{ccc}
 \sqrt{3}y_{e}\upsilon &  0 &  0 \\
 0 &  \sqrt{3}y_{\mu}\upsilon &  0 \\
 0 &  0 &  \sqrt{3}y_{\tau}\upsilon
 \end{array}\right)}~,~~~~~~~~~~{\rm with}~~
 U_{\omega}=\frac{1}{\sqrt{3}}{\left(\begin{array}{ccc}
 1 &  1 &  1 \\
 1 &  \omega &  \omega^{2} \\
 1 &  \omega^{2} &  \omega
 \end{array}\right)}
 \label{Yl2}
 \end{eqnarray}
 which indicates that the left-diagonalization matrices $V_{L}$ for the charged-lepton sector is identical as $U_{\omega}$ if their mass matrix can be diagonalized as $m^{\rm diag}_{f}=V^{\dag}_{L}m_{f}U_{R}=\upsilon{\rm Diag.}(y_{e}, y_{\mu}, y_{\tau})={\rm Diag.}(m_{e}, m_{\mu}, m_{\tau})$, and $U_R$ is the unit matrix.

Taking the scale of $A_{4}\times Z_{4}$ symmetry breaking to be above the electroweak scale in our scenario, that is, $\langle\chi\rangle, \langle\vartheta\rangle, \langle\psi\rangle > \langle\Phi^{0}\rangle$, and the required CP-violation can be supplied by the Yukawa interaction $\frac{f_{\nu}}{\Lambda}\bar{\ell}_{L}N_{R}\psi\eta$, being plus with $g_{\nu}\bar{\ell}_{L}N_{R}\eta$, the neutrino Yukawa coupling matrix is given as
 \begin{eqnarray}
 Y_{\nu}={\left(\begin{array}{ccc}
 g_{\nu} & f_{\nu}\frac{\upsilon_{\psi_{3}}}{\Lambda} &  f_{\nu}\frac{\upsilon_{\psi_{2}}}{\Lambda} \\
 f_{\nu}\frac{\upsilon_{\psi_{3}}}{\Lambda} & g_{\nu} &  f_{\nu}\frac{\upsilon_{\psi_{1}}}{\Lambda} \\
 f_{\nu}\frac{\upsilon_{\psi_{2}}}{\Lambda} & f_{\nu}\frac{\upsilon_{\psi_{1}}}{\Lambda} &  g_{\nu}
 \end{array}\right)}~,
 \label{YukawaCP}
 \end{eqnarray}
where $<\psi_{i}>=\upsilon_{\psi_{i}}~(i=1,2,3)$. The right-handed neutrino Majorana mass terms, being $M$ times the unity matrix plus being driven by $\langle\chi\rangle$, are given as
 \begin{eqnarray}
 M_{R}= {\left(\begin{array}{ccc}
 M &  \lambda_{\chi}\upsilon_{\chi_{3}} &  \lambda_{\chi}\upsilon_{\chi_{2}} \\
 \lambda_{\chi}\upsilon_{\chi_{3}} &  M &  \lambda_{\chi}\upsilon_{\chi_{1}} \\
 \lambda_{\chi}\upsilon_{\chi_{2}} &  \lambda_{\chi}\chi_{1} &  M
 \end{array}\right)}~,
 \label{MR1}
 \end{eqnarray}
where $<\chi_{i}>=\upsilon_{\chi_{i}}~(i=1,2,3)$ and $y_{R}<\vartheta>=M$.
If we assume the vacuum alignment of fields $\langle\chi_{i}\rangle$ and $\langle\psi_{i}\rangle$ can be chosen as follows
 \begin{eqnarray}
  \langle\chi_{1}\rangle&\equiv&\upsilon_{\chi}\neq0,~\langle\chi_{2}\rangle=\langle\chi_{3}\rangle=0~,\nonumber\\
  \langle\psi_{2}\rangle&\equiv&\upsilon_{\psi}\neq0,~\langle\psi_{1}\rangle=\langle\psi_{3}\rangle=0~,
 \label{subgroup}
 \end{eqnarray}
 $A_{4}\times Z_{4}$ symmetry is broken in such a way that while keeping $\mu-\tau$ symmetry in the right-handed Majorana mass term, Yukawa neutrino sector to be broken\footnote{It is equivalent to the way of $\mu-\tau$ symmetry breaking $\langle\psi_{3}\rangle\equiv\upsilon_{\psi}\neq0,~\langle\psi_{1}\rangle=\langle\psi_{2}\rangle=0$~.}.
The choice of VEV directions in Eq.~(\ref{subgroup}) and $\langle\Phi\rangle$ require a stable (or at least approximately stable) alignment of the fields $\chi,\vartheta,\psi$ and $\Phi$, which is displayed in Appendix in which we assume {\it ad hoc} constraints to realize the vacuum alignments. Then, we rewrite the right-handed Majorana neutrino mass term and neutrino Yukawa coupling matrix, which are given as
 \begin{eqnarray}
 M_{R}=M{\left(\begin{array}{ccc}
 1 &  0 &  0 \\
 0 &  1 &  \kappa e^{i\xi} \\
 0 &  \kappa e^{i\xi} &  1
 \end{array}\right)}~,~~~~Y_{\nu}=g_{\nu}{\left(\begin{array}{ccc}
 1 & 0 &  xe^{i\phi} \\
 0 & 1 &  0 \\
 xe^{i\phi} & 0 &  1
 \end{array}\right)}~,
 \label{MR2}
 \end{eqnarray}
where $\kappa=|\lambda_{\chi}|\upsilon_{\chi}/M$ and $x\equiv \upsilon_{\psi}|f_{\nu}|/(\Lambda g_{\nu})$. As will be shown later, the size of $x$ is restricted by the unknown mixing angle $\theta_{13}$.
Diagonalizing $M_{R}$ in order to go into the physical basis (mass basis) of the right-handed neutrino, the diagonalization of $M_{R}$ is given as
 \begin{eqnarray}
  M^{d}_{R} &=& V^{\dag}_{R}M_{R}V^{\ast}_{R}=M{\rm Diag.}(a, 1, b)~,
  \label{MR3}
 \end{eqnarray}
where $a=\sqrt{1+\kappa^{2}+2\kappa\cos\xi}, b=\sqrt{1+\kappa^{2}-2\kappa\cos\xi}$, with real and positive mass eigenvalues, $M_{1}=Ma, M_{2}=M, M_{3}=Mb$, and the diagonalizing matrix $V_{R}$ is
 \begin{eqnarray}
  V_{R}=\frac{1}{\sqrt{2}}{\left(\begin{array}{ccc}
  0  &  \sqrt{2}  &  0 \\
  -1 &  0  &  -1 \\
  -1 &  0  &  1
  \end{array}\right)}{\left(\begin{array}{ccc}
  e^{i\frac{\varphi_1}{2}}  &  0  &  0 \\
  0  &  1  &  0 \\
  0  &  0  &  e^{i\frac{\varphi_2}{2}}
  \end{array}\right)}~,
  \label{VR}
 \end{eqnarray}
with the phases
 \begin{eqnarray}
  \varphi_1=\tan^{-1}\Big(\frac{\kappa\sin\xi}{1+\kappa\cos\xi}\Big)~~~{\rm and}~~~\varphi_2=\tan^{-1}\Big(\frac{\kappa\sin\xi}{\kappa\cos\xi-1}\Big)~.
 \end{eqnarray}

In a basis where both charged lepton and heavy Majorana neutrino mass matrices are diagonal, the Yukawa interactions in Eq.~(\ref{lagrangian}) are replaced by
 \begin{eqnarray}
 {\cal L}_{\rm Yuk} &=& (\tilde{Y}_{\nu})_{j i} \bar{\ell}_{L j}\eta N_{i}+M^{d}_{R}\bar{N}_{i}(N_{i})^{c}+h.c
 \label{lagrangianA}
 \end{eqnarray}
where $M^{d}_{R}={\rm diag}(M_{1},M_{2},M_{3})$ and the couplings of $N_{i}$ with leptons and scalar $\eta$, $\tilde{Y}_{\nu}\equiv V^{\dag}_{L}Y_{\nu}V_{R}$, is given as
 \begin{eqnarray}
 \tilde{Y}_{\nu}=V^{e\dag}_{L}Y_{\nu}V_{R}= g_{\nu}{\left(\begin{array}{ccc}
 -\frac{2+e^{i\phi}x}{\sqrt{6}} & \frac{1+e^{i\phi}x}{\sqrt{3}} &  \frac{e^{i\phi}x}{\sqrt{6}} \\
 \frac{1-e^{i\phi}x}{\sqrt{6}} &  \frac{1+\omega e^{i\phi}x}{\sqrt{3}}  &  \frac{i\sqrt{3}+e^{i\phi}x}{\sqrt{6}} \\
 \frac{1-e^{i\phi}x}{\sqrt{6}} &  \frac{1-\omega e^{i\phi}x}{\sqrt{3}}  &  \frac{-i\sqrt{3}+e^{i\phi}x}{\sqrt{6}}
 \end{array}\right)}{\left(\begin{array}{ccc}
 e^{i\frac{\varphi_1}{2}}  &  0  &  0 \\
 0  &  1  &  0 \\
 0  &  0  &  e^{i\frac{\varphi_2}{2}}
 \end{array}\right)}~.
\label{Knu}
 \end{eqnarray}
Concerned with CP violation, we notice that the CP phases $\varphi_1,\varphi_2$ coming from $M_{R}$ as well as the CP phase $\phi$ from $Y_{\nu}$ obviously take part in low-energy CP violation, as you can see in Eq.~(\ref{meff3}). On the other hand, leptogenesis is associated with both $\tilde{Y}_{\nu}$ itself and the combination of neutrino Dirac Yukawa coupling matrix, $H\equiv\tilde{Y}^{\dag}_{\nu}\tilde{Y}_{\nu}= V^{\dag}_{R}Y^{\dag}_{\nu}Y_{\nu}V_{R}$, which is given as
 \begin{eqnarray}
 {\rm Im}\{H_{ij}(\tilde{Y}_{\nu})^{\ast}_{\alpha i}(\tilde{Y}_{\nu})_{\alpha j}\},~{\rm with}~
   H=g^{2}_{\nu}\left(\begin{array}{ccc}
  1+\frac{x^{2}}{2} & -\sqrt{2}xe^{-i\frac{\varphi_1}{2}}\cos\phi & -\frac{x^{2}}{2}e^{i\frac{\varphi_2-\varphi_1}{2}} \\
  -\sqrt{2}xe^{i\frac{\varphi_1}{2}}\cos\phi & 1+x^{2} & \sqrt{2}xe^{i\frac{\varphi_2}{2}}\cos\phi \\
  -\frac{x^{2}}{2}e^{i\frac{\varphi_1-\varphi_2}{2}} & \sqrt{2}xe^{-i\frac{\varphi_2}{2}}\cos\phi & 1+\frac{x^{2}}{2}
  \end{array}\right).
 \label{YnuYnu}
 \end{eqnarray}
where $\alpha=e,\mu,\tau$, which implies that both CP phases in $M_{R}$ and $Y_{\nu}$
take part in leptogenesis.

\section{neutrino mass matrix}

\begin{figure}[hbt]
\begin{picture}(360,100)(0,0)
\ArrowLine(90,10)(130,10)
\ArrowLine(180,10)(130,10)
\ArrowLine(180,10)(230,10)
\ArrowLine(270,10)(230,10)
\DashArrowLine(155,85)(180,60)3
\DashArrowLine(205,85)(180,60)3
\DashArrowArc(180,10)(50,90,180)3
\DashArrowArcn(180,10)(50,90,0)3
\Text(110,0)[]{$\nu_\alpha$}
\Text(250,0)[]{$\nu_\beta$}
\Text(180,0)[]{$N_i$}
\Text(135,50)[]{$\eta^0$}
\Text(230,50)[]{$\eta^0$}
\Text(150,90)[]{$\Phi^{0}$}
\Text(217,90)[]{$\Phi^{0}$}
\end{picture}
\caption{\label{Fig1} One-loop generation of light neutrino masses.}
\end{figure}
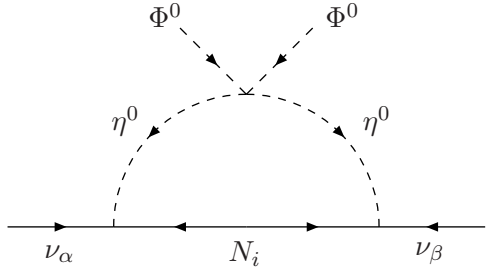
Due to the $Z_{2}$ symmetry, we can not get the neutrino Dirac masses and therefore the usual seesaw does not operate any more. However, similar to~\cite{Ma:2006fn} the light neutrino mass matrix can be generated through one-loop diagram showed in Fig.~\ref{Fig1} with the quadratic scalar interactions, i.e $\lambda_{\Phi\eta}(\Phi^{\dag}\eta)(\Phi^{\dag}\eta)$. After electroweak symmetry breaking, i.e. $\langle\Phi^{0}\rangle=(\upsilon,\upsilon,\upsilon)$, in a charged lepton mass matrix is diagonal, the flavor neutrino masses can be written as
 \begin{eqnarray}
  (m_{\nu})_{\alpha\beta}=\sum_{i}\frac{\tilde{Y}_{\nu\alpha i}\tilde{Y}_{\nu\beta i}}{M_{i}}F\Big(\frac{M^{2}_{i}}{\bar{m}^{2}_{\eta}}\Big)~,
 \label{radseesaw1}
 \end{eqnarray}
where $F(z_{i})=\frac{\Delta m^{2}_{\eta}}{16\pi^{2}}\big(\frac{z_{i}}{1-z_{i}}\big)\big[1+\frac{z_{i}\ln z_{i}}{1-z_{i}}\big]$, $\Delta m^{2}_{\eta}\equiv|m^{2}_{R}-m^{2}_{I}|=\mathcal{O}(\lambda_{\Phi\eta})\upsilon^{2}$ and $z_{i}=\frac{M^{2}_{i}}{\bar{m}^{2}_{\eta}}$, if $m_{R}(m_{I}) $ is the mass of $\eta^{0}_{R}(\eta^{0}_{I})$ and $m^{2}_{R(I)}=\bar{m}^{2}_{\eta}\pm\Delta m^{2}_{\eta}/2$\footnote{Actually, from potential lagrangian we can fully express the scalar masses $m_{\eta^{\pm}}, m_{R}, m_{I}$ and $\bar{m}_{\eta}$. However, for simplicity, these are expressed in terms of a relevant potential term.}.
 In our scenario, if we assumed $M^{2}_{i}\gtrsim \bar{m}^{2}_{\eta}$, so the lightest $Z_{2}$-odd neutral particle of $\eta$ is stable, the above formula Eq.~(\ref{radseesaw1}) can be written as,
 \begin{eqnarray}
  m_{\nu} = \frac{\Delta m^{2}_{\eta}}{16\pi^{2}}\tilde{Y}_{\nu}M^{-1}_{R}\tilde{Y}^{T}_{\nu}~,
 \label{radseesaw3}
 \end{eqnarray}
where $M_{R}={\rm Diag}(M_{r1},M_{r2},M_{r3})$ and $M_{ri}$ can be simplified as,
 \begin{eqnarray}
 M_{ri}\simeq\left\{\begin{array}{ll}
    2M_{i}, & \hbox{for $z_{i}\rightarrow1$} \\
    M_{i}\big[\ln z_{i}-1\big]^{-1} , & \hbox{for $z_{i}\gg1$.}
  \end{array}\right.
 \end{eqnarray}

{\bf How could we obtain the tri-bimaximal mixing matrix ?}
In the limit of $x\rightarrow0$ in Eq.~(\ref{MR2}), the light neutrino mass matrix in a basis where the charged lepton and heavy Majorana neutrino mass matrices are diagonal, that is, $Y_{\nu}$ is replaced by $\tilde{Y}_{\nu}\equiv V^{\dag}_{L}Y_{\nu}V_{R}$, can be obtained as follows
 \begin{eqnarray}
  m_{\nu} = \frac{m_{0}}{3}{\left(\begin{array}{ccc}
  \frac{2\omega_{1}e^{i\varphi_1}}{a}+\omega_{2} & \omega_{2}-\frac{\omega_{1}e^{i\varphi_1}}{a} &  \omega_{2}-\frac{\omega_{1}e^{i\varphi_1}}{a}  \\
  \omega_{2}-\frac{\omega_{1}e^{i\varphi_1}}{a}  & \frac{-3\omega_{3}e^{i\varphi_2}}{2b}+\frac{\omega_{1}e^{i\varphi_1}}{2a}+\omega_{2} &  \frac{3\omega_{3}e^{i\varphi_2}}{2b}+\frac{\omega_{1}e^{i\varphi_1}}{2a}+\omega_{2} \\
  \omega_{2}-\frac{\omega_{1}e^{i\varphi_1}}{a}  & \frac{3\omega_{3}e^{i\varphi_2}}{2b}+\frac{\omega_{1}e^{i\varphi_1}}{2a}+\omega_{2} &  \frac{-3\omega_{3}e^{i\varphi_2}}{2b}+\frac{\omega_{1}e^{i\varphi_1}}{2a}+\omega_{2}
 \end{array}\right)}~,
 \label{matrix1}
 \end{eqnarray}
where $\omega_{i}=\frac{z_{i}}{1-z_{i}}[1+\frac{z_{i}\ln z_{i}}{1-z_{i}}]$. And the overall scale of neutrino mass matrix $m_{0}$ is given as
 \begin{eqnarray}
  m_{0}=\frac{\Delta m^{2}_{\eta}}{16\pi^{2}}\frac{g^{2}_{\nu}}{M}~.
  \label{overall}
 \end{eqnarray}
This mass matrix can be diagonalized by the so-called tribimaximal mixing matrix $U_{TB}$ with mixing angles given in Eq.~(\ref{tibi1}),
 \begin{eqnarray}
  m_{\nu} =U_{\rm TB}{\rm Diag.}(m_{1},m_{2},m_{3})U^{T}_{\rm TB}=m_{0}U_{\rm TB}P_{\nu}{\rm Diag.}\Big(\frac{\omega_{1}}{a}, \omega_{2}, \frac{\omega_{3}}{b}\Big)P^{T}_{\nu}U^{T}_{\rm TB}~,
 \label{matrix2}
 \end{eqnarray}
we denote $m_{i}$ with $i=1-3$ which are the eigenvalues of $m_{\nu}$, and the matrices $U_{\rm TB}$ and $P_{\nu}$ are
 \begin{eqnarray}
 U_{\rm TB}={\left(\begin{array}{ccc}
  -\sqrt{\frac{2}{3}} &  \sqrt{\frac{1}{3}} &  0 \\
  \sqrt{\frac{1}{6}} &  \sqrt{\frac{1}{3}} &  \frac{1}{\sqrt{2}} \\
  \sqrt{\frac{1}{6}} &  \sqrt{\frac{1}{3}} &  \frac{-1}{\sqrt{2}}
 \end{array}\right)}~,~~~P_{\nu}={\left(\begin{array}{ccc}
  e^{i\frac{\varphi_1}{2}} & 0 & 0 \\
  0 & 1 & 0 \\
  0 & 0 & e^{i\frac{\varphi_2+\pi}{2}}
 \end{array}\right)}~.
 \label{TB}
 \end{eqnarray}

\subsection{Deviation from Tri-Bimaximal}
In order to achieve the deviation from the TBM matrix in neutrino sector we need to break $\mu-\tau$ symmetry, that is, $x\neq0$ in matrix $Y_{\nu}$ (see Eq.(\ref{MR2})) where the VEVs are aligned in Eq.(\ref{subgroup}). Thus the mass matrix of neutrinos can be written as
 \begin{eqnarray}
  m_{\rm eff} =
  m_{0}U_{\rm TB}P_{\nu}{\left(\begin{array}{ccc}
  \frac{\omega_{1}}{a}+\frac{e^{2i\phi}x^{2}\omega_{2}}{2} & -\frac{xe^{i\phi}}{\sqrt{2}}(\frac{\omega_{1}}{a}+\omega_{2}) &  -\frac{e^{2i\phi}x^{2}\omega_{2}}{2} \\
  -\frac{xe^{i\phi}}{\sqrt{2}}(\frac{\omega_{1}}{a}+\omega_{2}) & \omega_{2}+\frac{x^{2}e^{2i\phi}}{2}(\frac{\omega_{1}}{a}+\frac{\omega_{3}}{b}) &  \frac{xe^{i\phi}}{\sqrt{2}}(\frac{\omega_{3}}{b}+\omega_{2}) \\
  -\frac{e^{2i\phi}x^{2}\omega_{2}}{2} & \frac{xe^{i\phi}}{\sqrt{2}}(\frac{\omega_{3}}{b}+\omega_{2}) &  \frac{\omega_{3}}{b}+\frac{e^{2i\phi}x^{2}\omega_{2}}{2}
 \end{array}\right)}P^{T}_{\nu}U^{T}_{\rm TB}~,
  \label{meff3}
 \end{eqnarray}
which represents that $\mu-\tau$ symmetry is broken by $x$, and can not be diagonalized by $U_{\rm TB}$ in Eq.~(\ref{TB}).
To diagonalize the above matrix Eq.~(\ref{meff3}), if we consider $m_{\rm eff}m^{\dag}_{\rm eff}$ one can obtain the masses and the mixing angles.
For numerical purpose, we consider the case of $\varphi_{1,2}= 0$ without a loss of generality. Then, the light neutrino masses are given, up to first order of $x$, as
 \begin{eqnarray}
  |m_{1}|^{2} &\simeq& m^{2}_{0}\Big\{\frac{\omega^{2}_{1}}{a^{2}}-\frac{x}{4}\Big(\omega_{2}+\frac{\omega_{1}}{a}\Big)^{2}\cos\phi\Big\}~,\nonumber\\
  |m_{2}|^{2} &\simeq& m^{2}_{0}\Big\{\omega^{2}_{2}+\frac{x}{4}\Big(\omega_{2}+\frac{\omega_{1}}{a}\Big)^{2}\cos\phi\Big\}~, \nonumber\\
  |m_{3}|^{2} &\simeq& m^{2}_{0}\frac{\omega^{2}_{3}}{b^{2}}~.
  \label{eigenvalues}
 \end{eqnarray}
And the deviation from maximality of atmospheric neutrino mixing angle comes out as
 \begin{eqnarray}
  \theta_{23}+\frac{\pi}{4} \simeq \frac{6(\omega^{2}_{2}-\frac{\omega^{2}_{1}}{a^{2}})\sin\phi}{3(\omega^{2}_{2}-\frac{\omega^{2}_{1}}{a^{2}})\sin\phi+\sqrt{3}(\omega_{2}+\frac{\omega_{3}}{b})^{2}\cos\phi}~,
  \label{theta23}
 \end{eqnarray}
in which if the value of the parameter $\kappa$ is given by heavy neutrino mass ordering, the deviation from the maximality of atmospheric mixing angle can be determined only by the parameter $\phi$. From Eq.~(\ref{theta23}) we know that the values of $\phi$ at $\pi/2, 3\pi/2$ are not allowed by the experimental bounds of $\theta_{23}$. The unknown mixing angle $\theta_{13}$ and Dirac phase $\delta_{\rm CP}$ of $U_{\rm PMNS}$ can be obtained approximately, for $x\ll1$, by
 \begin{eqnarray}
  \theta_{13}&\simeq& -x\sqrt{\frac{3}{2}}\frac{(\frac{\omega^{2}_{3}}{b^{2}}-\omega^{2}_{2})\sin\phi\cos\delta_{CP}+(\omega_{2}+\frac{\omega_{3}}{b})^{2}\cos\phi\sin\delta_{CP}}{3\frac{\omega^{2}_{3}}{b^{2}}-\omega^{2}_{2}-\frac{\omega^{2}_{1}}{a^{2}}}~,\nonumber\\
  \delta_{CP} &\simeq& \tan^{-1}\Big(\frac{\omega_{3}+b\omega_{2}}{\omega_{3}-b\omega_{2}}\cot\phi\Big)~,
  \label{theta13}
 \end{eqnarray}
which indicates that $\theta_{13}$ is closely proportional to the size of $x$ and also related with $\phi$. From Eq.~(\ref{theta23}) and Eq.~(\ref{theta13}), we see that the deviation of $\theta_{23}$ is linked to $\theta_{13}$ through phase $\phi$, not through the parameter $x$.
Also, depending on the range of the phase $\phi$ we can expect the behavior of mixing angles $\theta_{23}, \delta_{CP}$ and $\theta_{13}$. Especially, Table-\ref{behavior} shows that $\theta_{13}$ is not allowed in the range of $\phi=0-\pi$, as well as how the mixing angle $\theta_{23}$ and the CP-phase $\delta_{\rm CP}$ behave depending on the parameter $\phi$, which will be shown in Fig.~\ref{Fig2} and Fig.~\ref{Fig4}.
\begin{table}[th]
\begin{center}
\begin{tabular}{|c|c|c|c|c|} \hline
$\phi$ & $0\sim \pi/2$ & $\pi/2\sim\pi$ & $\pi\sim3\pi/2$ & $3\pi/2\sim2\pi$  \\ \hline \hline
$\theta_{23}+\frac{\pi}{4}$  &  $\theta_{23}<\pi/4$   &  $\theta_{23}>\pi/4$  &  $\theta_{23}<\pi/4$  &  $\theta_{23}>\pi/4$   \\ \hline
$\delta_{\rm CP}$ & $+$ & $-$  & $+$  & $-$  \\ \hline
$\theta_{13}$ & $-$ & $-$  & $+$ & $+$   \\ \hline
\end{tabular}
\caption{\label{behavior}  The behavior of mixings $\theta_{23}, \delta_{CP}$ and $\theta_{13}$ depending on the range of parameter $\phi$.}
\end{center}
\end{table}
And, solar neutrino mixing is governed by
 \begin{eqnarray}
  \tan2\theta_{12} \simeq 2\sqrt{2}\frac{\omega^{2}_{2}-\frac{\omega^{2}_{1}}{a^{2}}+\frac{x}{2}(\frac{\omega_{1}}{a}+\omega_{2})^{2}\cos\phi+\frac{x^{2}}{2}\{(\frac{\omega_{3}}{b}+\omega_{2})^{2}+2\omega_{2}\frac{\omega_{3}}{b}\cos2\phi\}}{\omega^{2}_{2}-\frac{\omega^{2}_{1}}{a^{2}}-4x(\frac{\omega_{1}}{a}+\omega_{2})^{2}\cos\phi+\frac{x^{2}}{2}\{(\frac{\omega_{3}}{b}+\omega_{2})^{2}+2\omega_{2}\frac{\omega_{3}}{b}\cos2\phi\}}
  \label{theta12}
 \end{eqnarray}
 which for $x=0$ agrees with the result of tri-bimaximal, i.e. $\tan2\theta_{12}=2\sqrt{2}$. Note here that in Eq.~(\ref{theta12}) the condition
 \begin{eqnarray}
  \omega^{2}_{2}-\frac{\omega^{2}_{1}}{a^{2}}+\frac{x^{2}}{2}\{(\frac{\omega_{3}}{b}+\omega_{2})^{2}+2\omega_{2}\frac{\omega_{3}}{b}\cos2\phi\}\gg|x(\frac{\omega_{1}}{a}+\omega_{2})^{2}\cos\phi|
 \end{eqnarray}
  should be satisfied, in order for $\theta_{12}$ to be lie in the experimental bounds in Table-\ref{tab:data}. Interesting points are that the deviation of $\theta_{12}$ from tri-bimaximal is closely related with $\theta_{13}$ through the parameters $x$ and $\phi$, and the deviation of $\theta_{23}$ from maximality is governed by the phase $\phi$ which is related with $\delta_{CP}$ in Eq.~(\ref{theta13}), if the parameter $\kappa$ is determined by heavy neutrino mass ordering.

\begin{figure}[ht]
  \centering
    \includegraphics[width=0.6\textwidth]{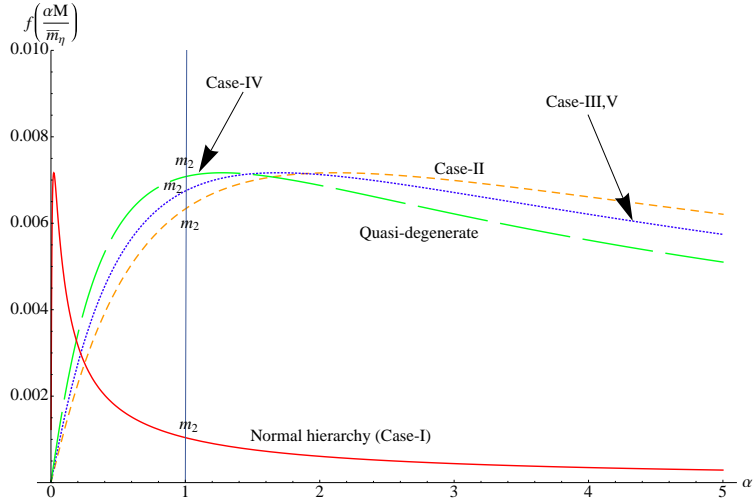}
  \caption{Neutrino mass ordering in different values of $\frac{M}{\bar{m}_{\eta}} = 1.001,1.25,1.6,100$, which correspond to {\bf Case-II}, {\bf Case-III,V}, {\bf Case-IV}, {\bf Case-I},  respectively. }
  \label{fig:nu-mass}
\end{figure}

\begin{figure}[ht]
\hspace*{-2cm}
\begin{minipage}[t]{6.0cm}
\epsfig{figure=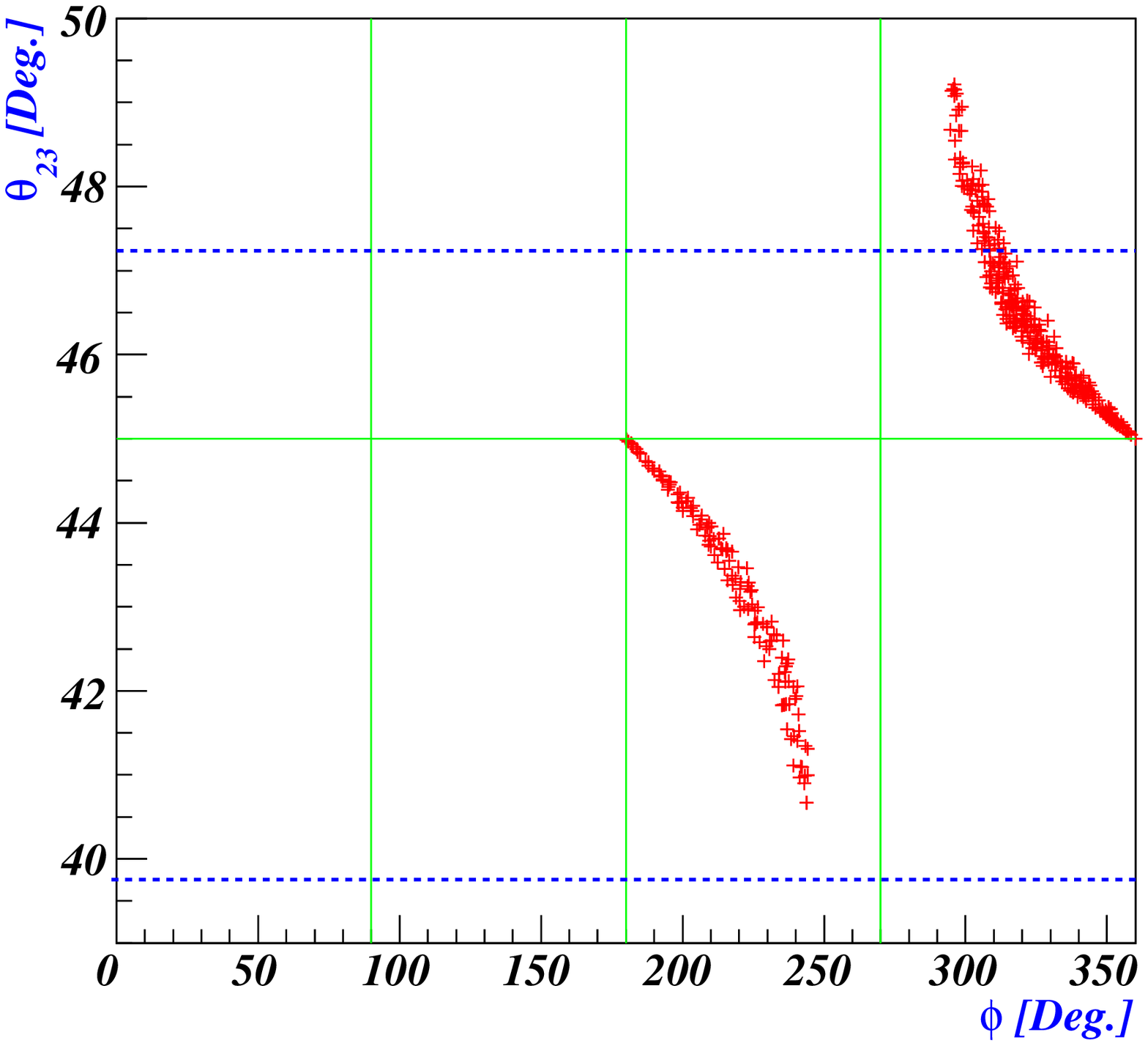,width=6.5cm,angle=0}
\end{minipage}
\hspace*{2.0cm}
\begin{minipage}[t]{6.0cm}
\epsfig{figure=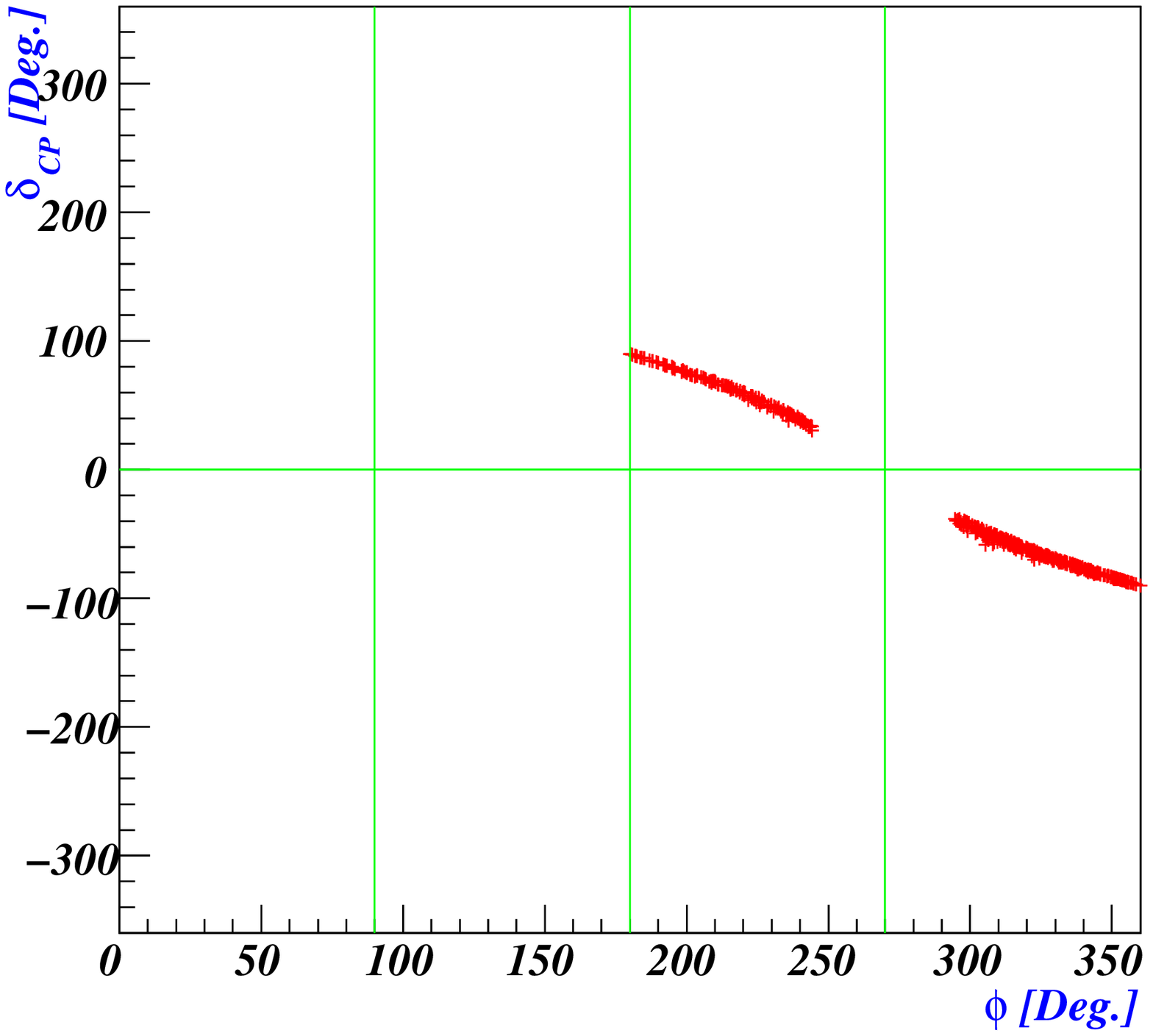,width=6.5cm,angle=0}
\end{minipage}
\vspace*{-1.0cm} \hspace*{-2cm}
\begin{minipage}[t]{6.0cm}
\epsfig{figure=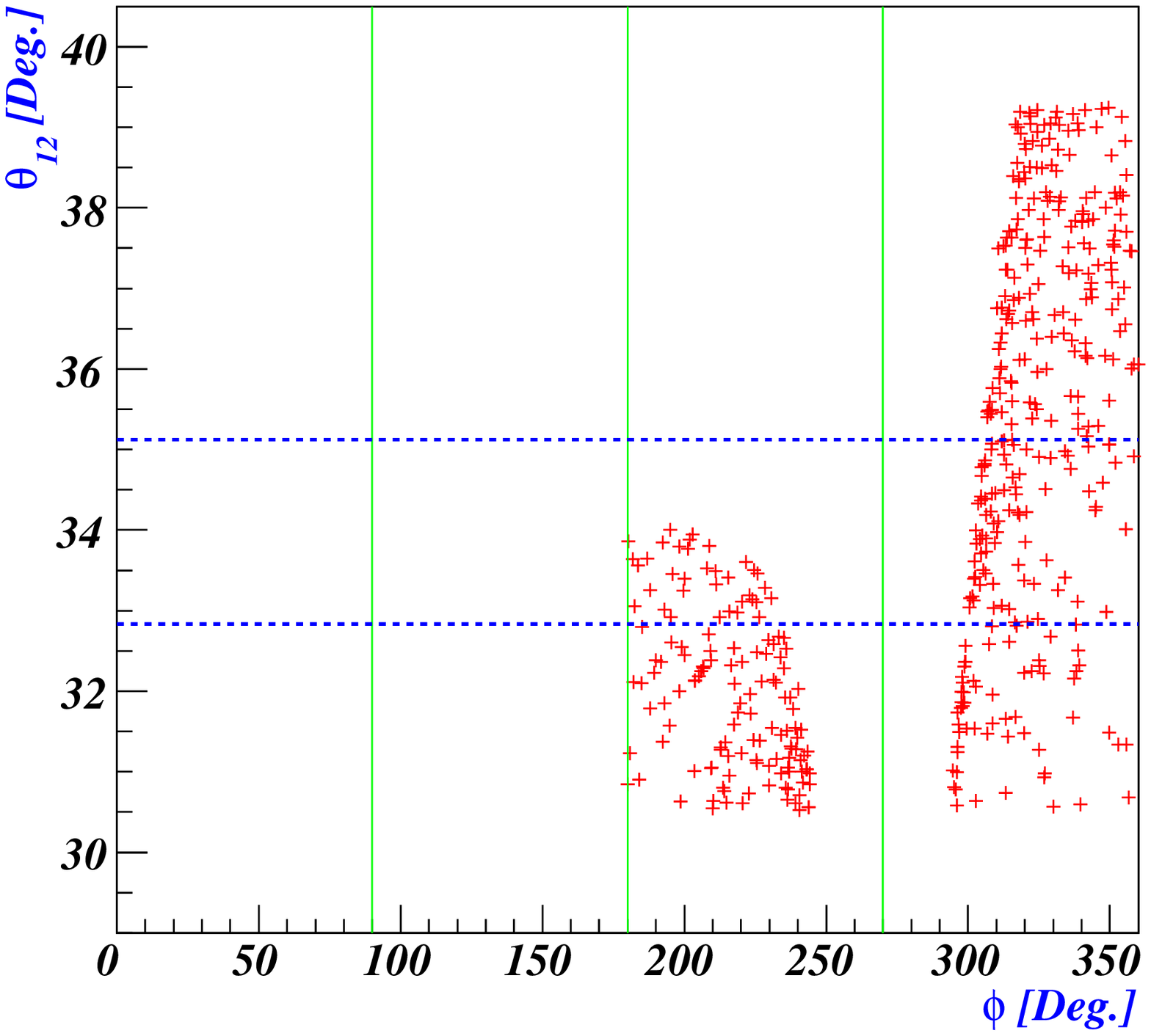,width=6.5cm,angle=0}
\end{minipage}
\hspace*{2.0cm}
\begin{minipage}[t]{6.0cm}
\epsfig{figure=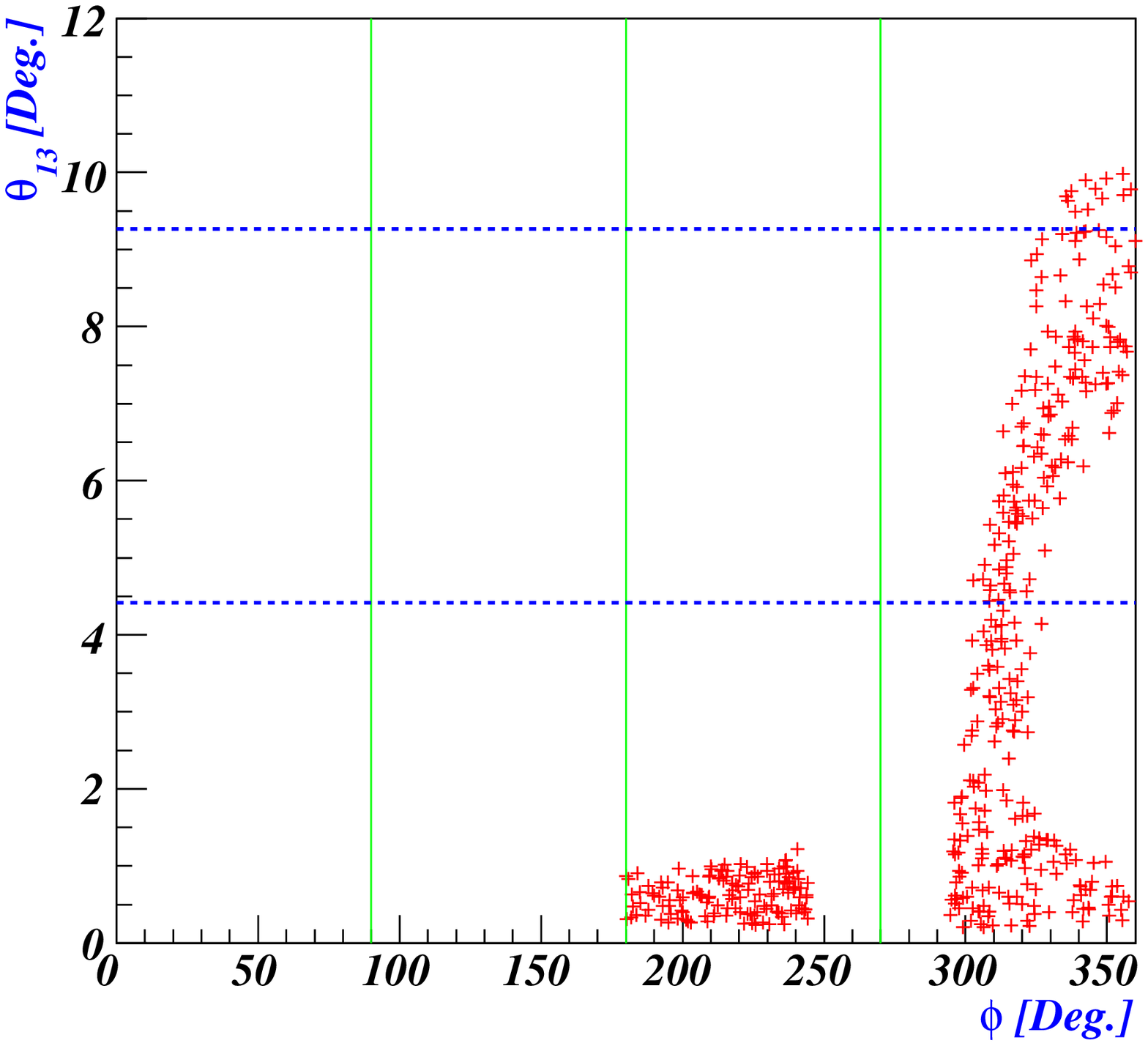,width=6.5cm,angle=0}
\end{minipage}
\caption{\label{Fig2}(Upper-panel:) Left-figure represents that the atmospheric mixing angle $\theta_{23}$ over the phase $\phi$. Right-figure represents the relation between the Dirac-CP phase $\delta_{CP}$ and the phase $\phi$. Here the horizontal dotted lines represent the experimental lower and upper bounds in $1\sigma$ of the mixing angle $\theta_{23}$. (Lower-panel:) Left-figure shows the mixing angle
$\theta_{12}$ as a function of the parameter $\phi$.  Right-figure shows the mixing angle $\theta_{13}$ as a function of the parameter $\phi$. Here the horizontal dotted lines represent the experimental upper and lower bound in $1\sigma$ of the mixing angles $\theta_{12}$ and $\theta_{13}$.}
\end{figure}

Because of the observed hierarchy $|\Delta m^{2}_{32}|\gg\Delta m^{2}_{21}$, and the requirement of MSW resonance for solar neutrinos, there are two possible neutrino mass spectrum: (i) $m_{1}<m_{2}<m_{3}$ (normal mass spectrum) which corresponds to $\frac{\omega_{1}}{a}<\omega_{2}<\frac{\omega_{3}}{b}$ and (ii) $m_{3}<m_{1}<m_{2}$ (inverted mass spectrum) which corresponds to $\frac{\omega_{3}}{b}<\frac{\omega_{1}}{a}<\omega_{2}$. Here we can approximate light neutrino masses in our model as $m_{i} \approx m_{0}(\frac{\omega_{1}}{a},\omega_{2},\frac{\omega_{3}}{b})$, by using Eqs.~(\ref{matrix1}), (\ref{overall}) we can express neutrino masses as,
\begin{eqnarray}
m_i &=& \frac{g^2_{\nu}m_{-}}{8\pi^2}\frac{\bar{m}_{\eta}}{M}\frac{1}{\alpha}\frac{\alpha^2M^2/\bar{m}^2_{\eta}}{1-\alpha^2M^2/\bar{m}^2_{\eta}}\Big[1 + \frac{\alpha^2M^2/\bar{m}^2_{\eta}\ln{\frac{\alpha^2M^2}{\bar{m}^2_{\eta}}}}{1 - \alpha^2M^2/\bar{m}^2_{\eta}}\Big] \nonumber \\
&=& g^2_{\nu}m_{-}f\Big(\alpha\frac{M}{\bar{m}_{\eta}}\Big),
\end{eqnarray}
where $\Delta m^{2}_{\eta}=m_{+}m_{-}$ and $m_+\simeq2\bar{m}_{\eta}$ are used, and $m_{\pm}$ denotes $|m_{R} \pm m_{I}|$. And $\alpha$ is a dummy index, $\alpha = a, 1, b$ refer to light neutrinos $m_1, m_2, m_3$ and also heavy neutrinos $M(a,1,b) = (M_{1},M_{2},M_{3})$ respectively.  Note that the overall scale of neutrino masses is determined by $g^2_{\nu}m_{-}$ as will be shown in Eq.~(\ref{mo1}), while the magnitudes of function $f(\alpha\frac{M}{\bar{m}_{\eta}})$ does not change a lot within the parameter region as we showed in Fig.~\ref{fig:nu-mass}. The neutrino spectrums are related to the ratio $\frac{M}{\bar{m}_{\eta}}$ and the value of $\alpha$, and $\alpha = 1$ corresponds to the mass of the second generation of light neutrinos $m_2$. The locations of $m_1$ and $m_3$ are determined by the values of $\alpha's$ (or $a,b$) which are defined in Eq.~(\ref{MR3}), and the constraints come from the solar and atmospheric mass-squared differences.
They are given by
 \begin{eqnarray}
  \Delta m^{2}_{21} &\equiv& m^{2}_{2}-m^{2}_{1} \simeq m^{2}_{0}\Big\{\omega^{2}_{2}-\frac{\omega^{2}_{1}}{a^{2}}+\frac{x}{2}\big(\omega_{2}+\frac{\omega_{1}}{a}\big)^{2}\cos\phi\Big\}~,
  \label{massSqd1}
 \end{eqnarray}
 \begin{eqnarray}
  \Delta m^{2}_{32}&\equiv& m^{2}_{3}-m^{2}_{2}\simeq m^{2}_{0}\Big\{\frac{\omega^{2}_{3}}{b^{2}}-\omega^{2}_{2}-\frac{x}{4}\big(\omega_{2}+\frac{\omega_{1}}{a}\big)^{2}\cos\phi\Big\}~,
  \label{massSqd2}
 \end{eqnarray}
in which, from the neutrino oscillation experiments we know that $\Delta m^{2}_{21}$ is positive and dictates $\omega_{2}>\omega_{1}/a$ with the second term being sufficiently small $x$ in Eq.~(\ref{massSqd1}). As will be shown later, in order for a leptogenesis to be successfully implemented at or around TeV scale in our scenario, we consider the case $M^{2}_{\rm lightest}\simeq \bar{m}^{2}_{\eta}$ where the $M_{\rm lightest}$ is the lightest of the heavy Majorana neutrino. Depending on the hierarchy of the heavy Majorana neutrino masses $M_{1}$, $M_{2}$ and $M_{3}$, the relative
size of the parameter $\kappa$ consistent with the possible mass ordering of light neutrinos and hierarchy of $\Delta m^{2}_{32}$ and $\Delta m^{2}_{21}$ can be classified as follows:
\begin{itemize}
  \item {\bf Case-I} $M_{1,2}\gg M_{3}$ ($a>1\gg b$ with $\xi=0$): this case corresponds to the normal hierarchical mass spectrum with $b\rightarrow0$ i.e. $\kappa\simeq1$. Using $\omega_{1}\simeq2\ln\frac{2}{b}-1$, $\omega_{2}\simeq-2\ln b-1$ and $\omega_{3}\simeq\frac{1}{2}$, the ratio of the mass squared differences defined by $R\equiv \frac{\Delta m^{2}_{21}}{|\Delta m^{2}_{32}|}$ which is around $R\simeq3\times10^{-2}$ for the best-fit values of the solar and atmospheric mass squared differences, which is given by
 \begin{eqnarray}
  R\approx b^{2}(4\omega^{2}_{2}-\omega^{2}_{1})~,
  \label{RM3}
 \end{eqnarray}
 where the equality roughly can be given under $1\gg x,b$. Note here that using the best-fit value of $R(\simeq3\times10^{-2})$ and Eq.~(\ref{RM3}) one can roughly determine the size of the parameter $b$, i.e $b\simeq0.01$.
  \item {\bf Case-II}  $M_{1}>M_{3}>M_{2}$ ($a>b>1$ with $\xi=0$): this corresponds to $\frac{\omega_{3}}{b}>\omega_{2}\gtrsim\frac{\omega_{1}}{a}$, the solution exists for $\frac{\omega_{1}}{a}$ going to $\omega_{2}\simeq\frac{1}{2}$ giving $a\simeq4.7$ and $b\simeq2.7$ with $\kappa\simeq3.7$ in numerical calculations, and gives a degenerate normal ordering of light neutrinos.
  \item {\bf Case-III} $M_{3}>M_{2}>M_{1}$ ($b>1\gtrsim a$ with $\xi=\pi$): this corresponds to a degenerate inverted ordering of light neutrinos giving $a\simeq1, b\simeq3$ with $\kappa\simeq2$ in numerical calculations.
  \item {\bf Case-IV} $M_{1}> M_{2}> M_{3}$ ($a>1>b$ with $\xi=0$): this case gives $a\simeq1.4$ and $b\approx0.6$ with $\kappa\simeq0.4$ in numerical calculations, which corresponds to $\omega_{2}\gtrsim\frac{\omega_{1}}{a}\gtrsim\frac{\omega_{3}}{b}$ indicating a degenerate inverted ordering of light neutrinos.
  \item {\bf Case-V} $M_{1}> M_{2}\gtrsim M_{3}$ ($a>1\gtrsim b$ with $\xi=0$): this case gives $a\simeq2.8, b\simeq0.8$ with $\kappa\simeq1.8$ in numerical calculations, which corresponds to $\omega_{2}\gtrsim\frac{\omega_{1}}{a}\gtrsim\frac{\omega_{3}}{b}$ indicating a degenerate inverted ordering of light neutrinos.
\end{itemize}
Note here that in our scenario the inverted hierarchical light neutrino mass spectrum is not allowed because the condition $\Delta m^{2}_{21}>0$ is not satisfied due to the mass ordering of heavy Majorana neutrinos Eq.~(\ref{MR3}) corresponding to light neutrino mass ordering.
In the expressions of Eqs.~(\ref{theta23}-\ref{massSqd2}), the values of parameters $\kappa$ (or $a,b$), $x,\phi$ can be determined from the analysis
described in above, whereas $g_{\nu}$ is arbitrary. However, since $m_{0}=\Delta m^{2}_{\eta}g^{2}_{\nu}/16\pi^{2}M$ as defined in Eq.~(\ref{overall}), the value of $g_{\nu}$ depends on the magnitude of $m_{-}$ in the case that $m_{0}$ is determined as
 \begin{eqnarray}
  m_{0}\simeq\frac{m_-bg^2_{\nu}}{8\pi^2} \simeq \left\{
            \begin{array}{ll}
              m_{3}b/\omega_{3}, & \hbox{Normal hierarchical mass spectrum~({\bf Case-I}),} \\
              m_{2}/\omega_{2}, & \hbox{Quasi-degenerate mass spectrum~({\bf Case-II$\sim$Case-V}).} \\
            \end{array}
          \right.
 \label{mo1}
 \end{eqnarray}
Since all new scalars $\eta^{\pm}, \eta^{0}_{R}, \eta^{0}_{I}$ carry a $Z_{2}$ odd quantum number and only couple to Higgs boson and electroweak gauge bosons of the standard model, they can be produced in pairs through the standard model gauge bosons $W^{\pm}, Z$ or $\gamma$.  Once produced, $\eta^{\pm}$ will decay into $\eta^{0}_{R,I}$ and a virtual $W^{\pm}$, then $\eta^0_{I}$ subsequently becomes $\eta^0_{R}$ + $Z$-boson, which will decay a quark-antiquark or lepton-antilepton pair. Here the mass hierarchy $m_{\eta^{\pm}}>m_{I}>m_{R}$ is assumed. That is, the stable $\eta^{0}_{R}$ appears as missing energy in the decays of $\eta^{\pm}\rightarrow\eta^{0}_{I}l^{\pm}\nu$ with the subsequent decay $\eta^{0}_{I}\rightarrow\eta^{0}_{R}l^{\pm}l^{\mp}$, which can be compared to the direct decay $\eta^{\pm}\rightarrow\eta^{0}_{R}l^{\pm}\nu$ to extract the masses of the respective particles. Therefore, if the signal of $m_{-}$ and $m_{+}$ in LHC are measured, i.e. $\bar{m}_{\eta}\simeq M_{\rm lightest}\gtrsim {\rm electroweak ~scale}$, the lightest of heavy Majorana neutrinos can be decided.

\subsection{Confronting with Low-energy neutrino data}
Before we discussing how to achieve leptogenesis in our scenario, we first examine if it is consistent with low energy neutrino data, especially being consistent with the recent analysis in $1\sigma$ giving $\theta_{13}>0$ \cite{bari}. As can be seen from Eqs.~(\ref{eigenvalues}-\ref{theta12}), three neutrino masses, three mixing angles and a CP phase are presented in terms of five independent parameters $m_{0}, \kappa ({\rm or}~a,b), x, \phi$. Note here that the values of parameter $\omega_{i} (i=1,2,3)$ and $a,b$ are determined independetly by the value of parameter $\kappa$. At present, we have five experimental results, which are taken as inputs in our numerical analysis given at $3\sigma$ by Table.~\ref{tab:data}.

Let us discuss the numerical results focussing on both hierarchical and degenerate light neutrino mass spectrum given in previous section, for example, {\bf Case-I} and {\bf Case-II}, respectively.
\subsubsection{Normal hierarchical light neutrino mass spectrum}
 In our numerical calculation of {\bf Case-I}, we first fix the value of heavy Majorana neutrino with lightest one being to be around TeV scale $M_{2}\equiv M=10^{5}{\rm GeV}\gg M_{3}\equiv bM\simeq 1{\rm TeV}$ and the value of $m_{0}$ with $m_{-}=100{\rm eV}$ in Eq.~(\ref{mo1}), then we impose the current experimental results on neutrino masses and mixings into the hermitian matrix $m^{\dag}_{\rm eff}m_{\rm eff}$ and varying all the parameter space $\{\kappa({\rm or}~a,b),\phi,x,g_{\nu}\}$:
 \begin{eqnarray}
  0.98\lesssim\kappa\lesssim1.02~,~~~~0\leq\phi\leq2\pi~,~~0.005\leq x<0.4~,~~~0.26\lesssim g_{\nu}\lesssim0.30~,
 \end{eqnarray}
where the parameter $g_{\nu}$ can be replaced by $m_{0}$ due to Eq.~(\ref{mo1}).
 As a result of the numerical analysis concerned with the mixing angle $\theta_{12}$ and $\theta_{13}$, we found that in the case of normal hierarchical light neutrino mass spectrum corresponding to $M_{3}\ll M_{1,2}$ the value of parameter $b$ is in the order of ${\cal O}(0.01)$, in turn which means that the second order of $x$ in Eq.~(\ref{theta12}) is also important to the contribution of $\theta_{12}$, allowing the values of $\theta_{13}$ to be lie in the experimental bounds in $1\sigma$.

\begin{figure}[t]
\hspace*{-2cm}
\begin{minipage}[t]{6.0cm}
\epsfig{figure=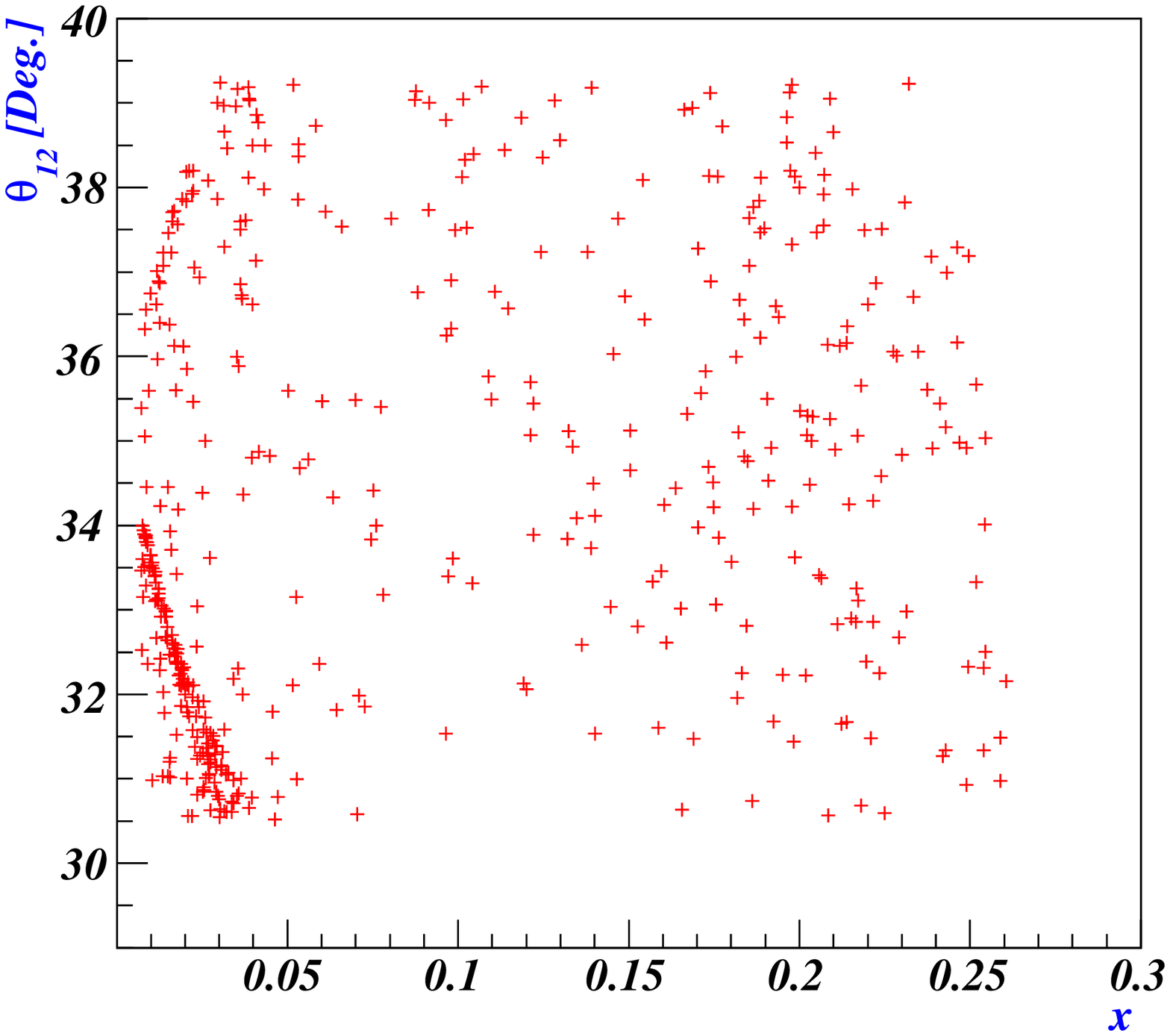,width=6.5cm,angle=0}
\end{minipage}
\hspace*{1.0cm}
\begin{minipage}[t]{6.0cm}
\epsfig{figure=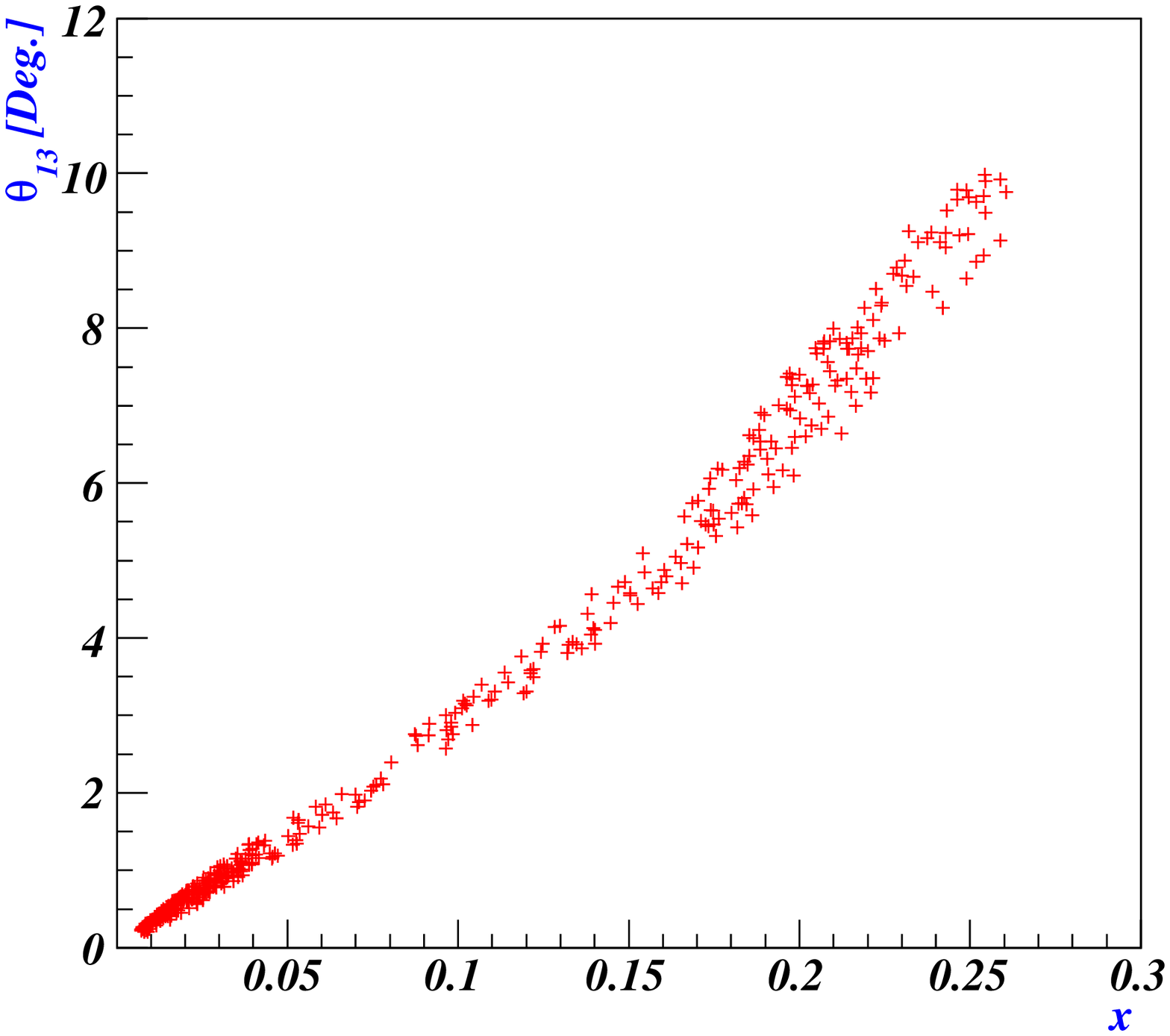,width=6.5cm,angle=0}
\end{minipage}
\caption{ \label{Fig3} Left-figure shows the mixing angle of
$\theta_{12}$ as a function of the parameter $x$. Right-figure shows
$\theta_{13}$ as a function of the parameter $x$.}
\end{figure}

Fig.~\ref{Fig2} shows how the mixing angles $\theta_{12}, \theta_{13}, \theta_{23}$ and $\delta_{\rm CP}$ in neutrino oscillation depend on the parameter $\phi$, which can be explained in the approximate analysis Eqs.~(\ref{theta23}-\ref{theta12}).
Fig.~\ref{Fig3} represents that how the mixing angles $\theta_{12}$ and $\theta_{13}$ depend on the parameter $x$, as can be seen in Eqs.~(\ref{theta13}-\ref{theta12}), in which especially the unknown mixing angle $\theta_{13}$ is very sensitive to the parameter $x$.

\subsubsection{Quasi-degenerate light neutrino mass spectrum}
On the other hand, in the case of degenerate light neutrino mass spectrum corresponding to {\bf Case-II}$\sim${\bf Case-V}, since the values of $b$ are not so small compared to the case of normal hierarchical mass spectrum, in a good approximation, the second order of $x$ in Eq.~(\ref{theta12}) can be safely neglected, and Eq.~(\ref{theta12}) can be simplified as
 \begin{eqnarray}
  \tan2\theta_{12} \simeq 2\sqrt{2}\frac{(\omega_{2}-\frac{\omega_{1}}{a})+\frac{x}{2}(\frac{\omega_{1}}{a}+\omega_{2})\cos\phi}{(\omega_{2}-\frac{\omega_{1}}{a})-4x(\frac{\omega_{1}}{a}+\omega_{2})\cos\phi}~,
  \label{theta12a}
 \end{eqnarray}
in which for $\theta_{12}$ to be lie in the range of experimental bounds the condition
 \begin{eqnarray}
  \omega_{2}-\frac{\omega_{1}}{a}\gg|x(\frac{\omega_{1}}{a}+\omega_{2})\cos\phi|
 \end{eqnarray}
is required, which means that for $\omega_{2}\approx\frac{\omega_{1}}{a}\lesssim\frac{\omega_{3}}{b}$ the values of $x\cos\phi$ should be very small.
Therefore, for degenerate light neutrino mass spectrum we can expect a very small value of $x$ and a very small $\theta_{13}$, except in the limit of $\phi\rightarrow3\pi/2$ (not in $\phi=3\pi/2$) in which the value of $x$ can be large and in turn a large $\theta_{13}<0.2$ can be expected.

For example, in our numerical calculation of {\bf Case-II}, we first fix the value of heavy Majorana neutrino with lightest one being to be TeV scale $M_{2}\equiv M=1{\rm TeV}$ and the value of $m_{0}$ with $m_{-}=100{\rm eV}$ in Eq.~(\ref{mo1}), then we impose the current experimental results on neutrino masses and mixings into the hermitian matrix $m^{\dag}_{\rm eff}m_{\rm eff}$ and varying all the parameter space $\{\kappa({\rm or}~a,b),\phi,x,g_{\nu}\}$:
 \begin{eqnarray}
  3.2\lesssim\kappa\lesssim3.8~,~~~~0\leq\phi\leq2\pi~,~~0.005\leq x<0.4~,~~~0.38\lesssim g_{\nu}\lesssim0.42~,
 \end{eqnarray}
where the parameter $g_{\nu}$ can be replaced by $m_{0}$ due to Eq.~(\ref{mo1}).

\begin{figure}[t]
\hspace*{-2cm}
\begin{minipage}[t]{6.0cm}
\epsfig{figure=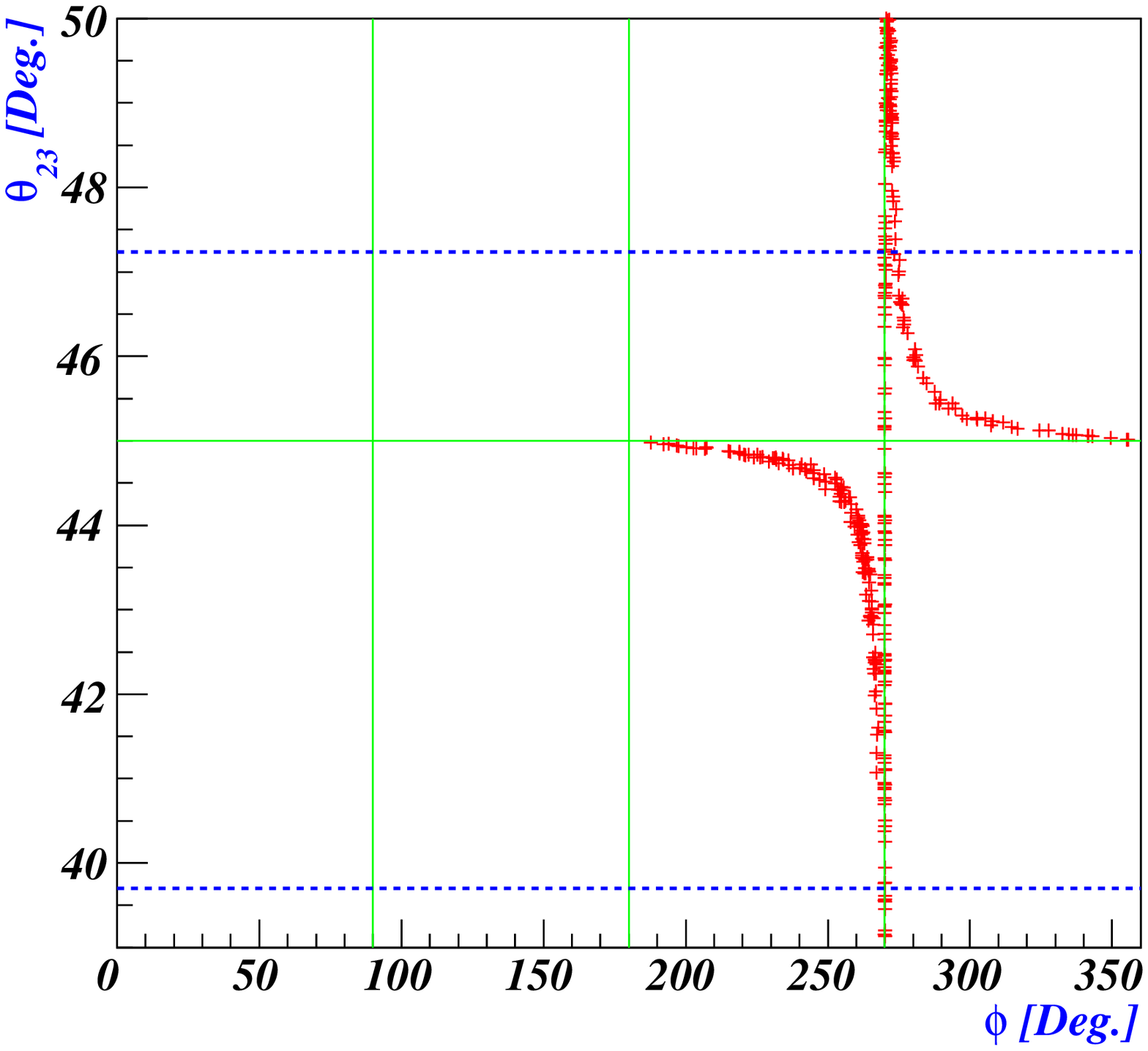,width=6.5cm,angle=0}
\end{minipage}
\hspace*{2.0cm}
\begin{minipage}[t]{6.0cm}
\epsfig{figure=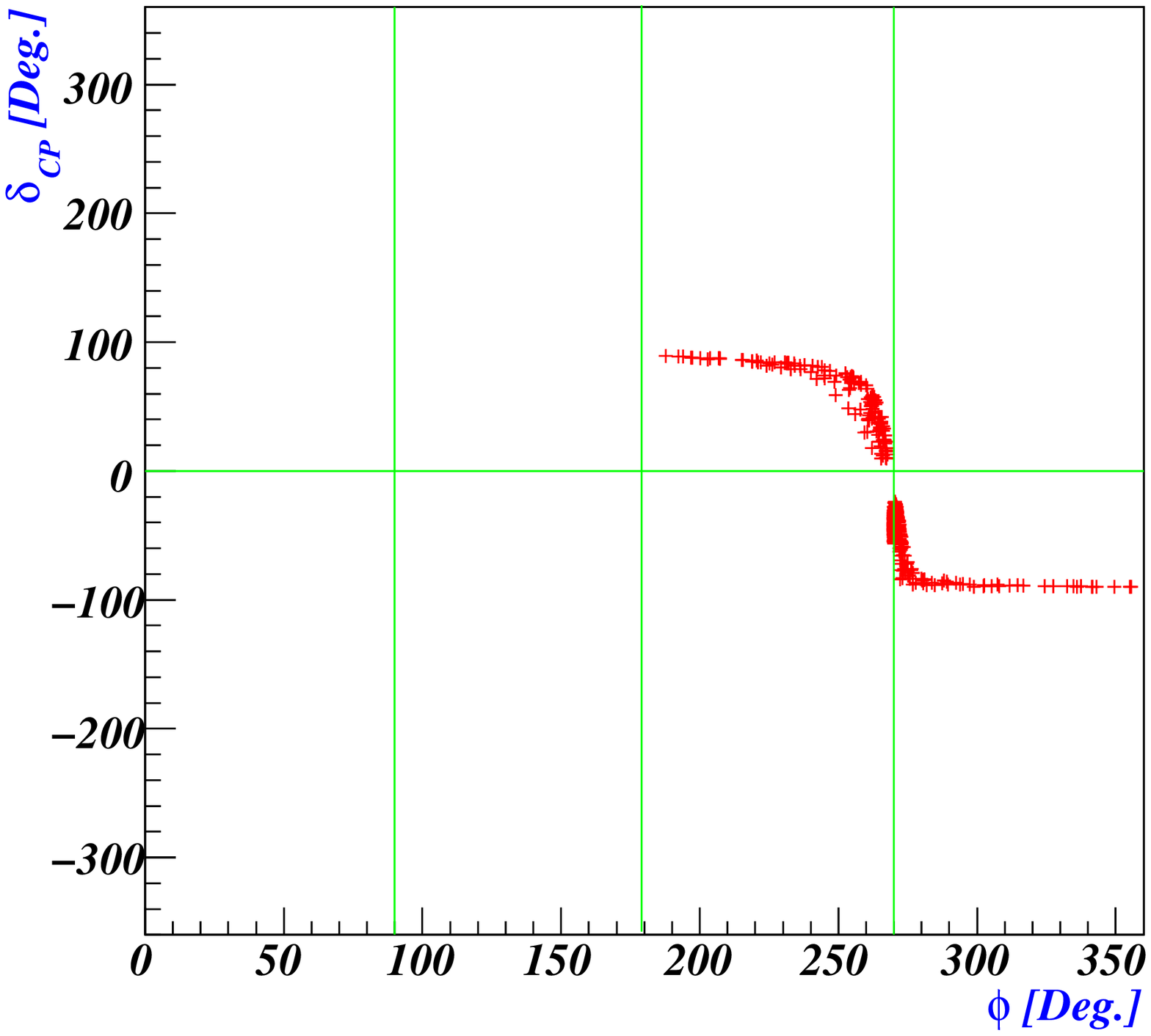,width=6.5cm,angle=0}
\end{minipage}
\vspace*{-1.0cm} \hspace*{-2cm}
\begin{minipage}[t]{6.0cm}
\epsfig{figure=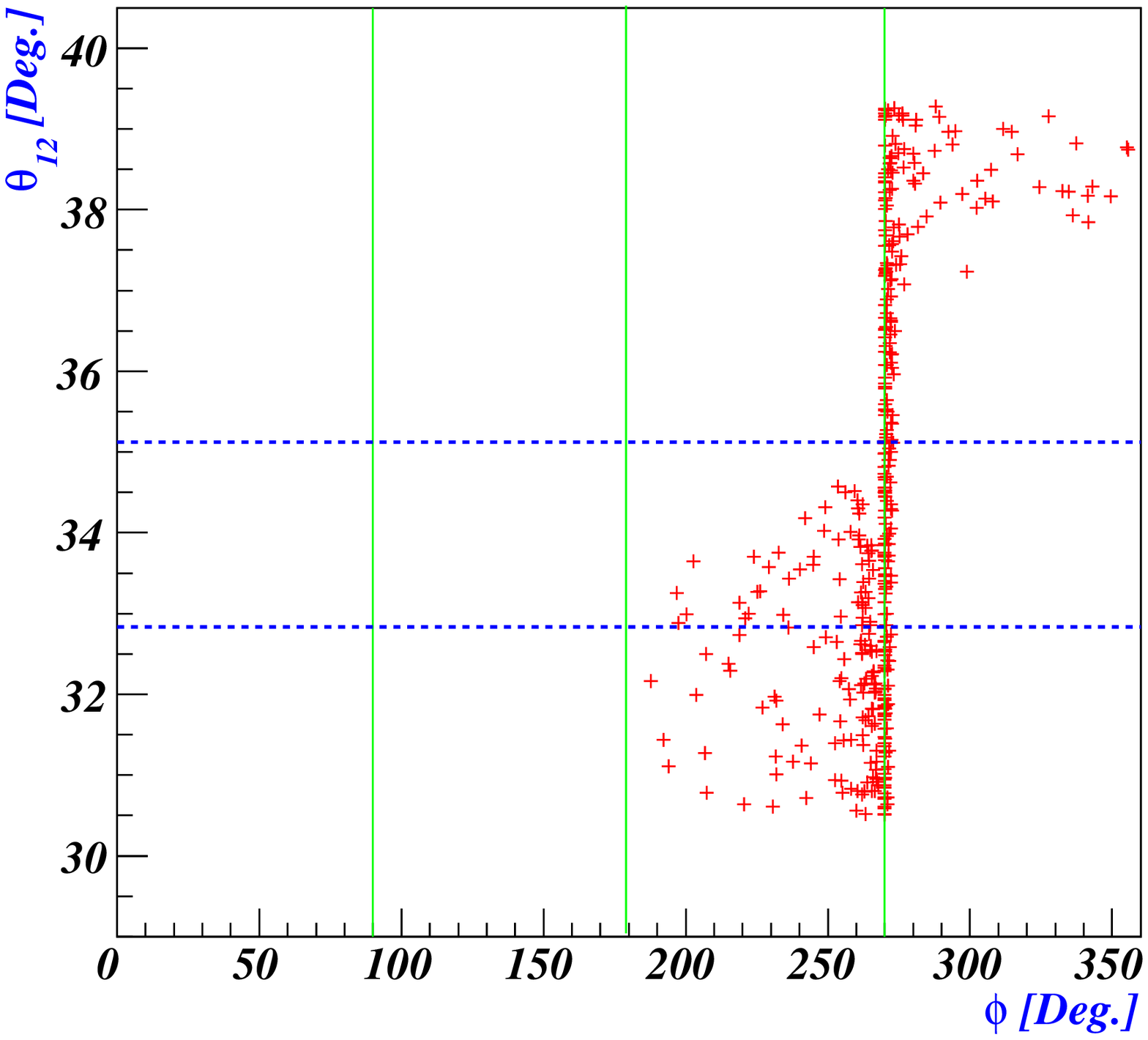,width=6.5cm,angle=0}
\end{minipage}
\hspace*{2.0cm}
\begin{minipage}[t]{6.0cm}
\epsfig{figure=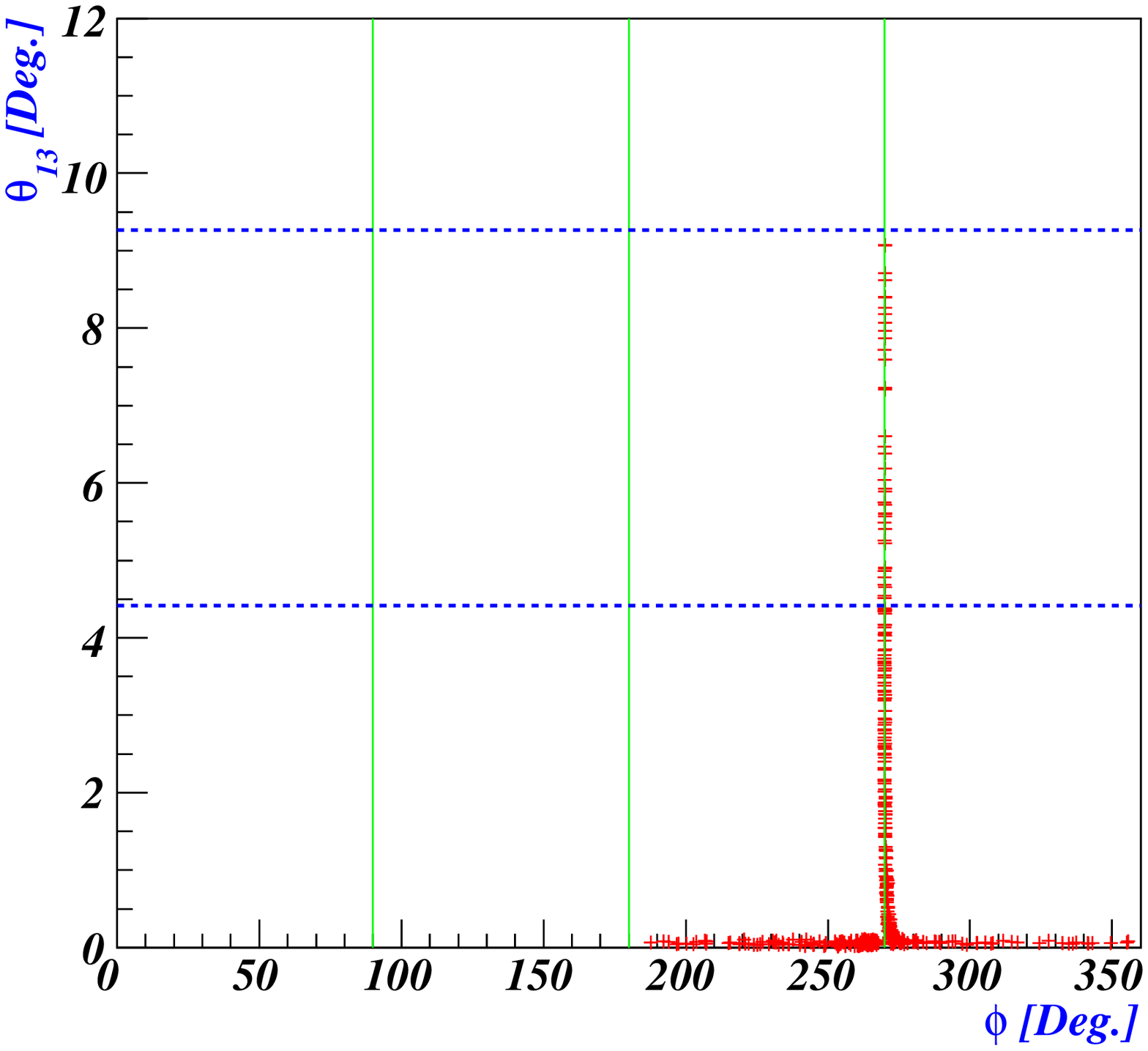,width=6.5cm,angle=0}
\end{minipage}
\caption{\label{Fig4} The same as Fig.\ref{Fig2}.}
\end{figure}

\begin{figure}[t]
\hspace*{-2cm}
\begin{minipage}[t]{6.0cm}
\epsfig{figure=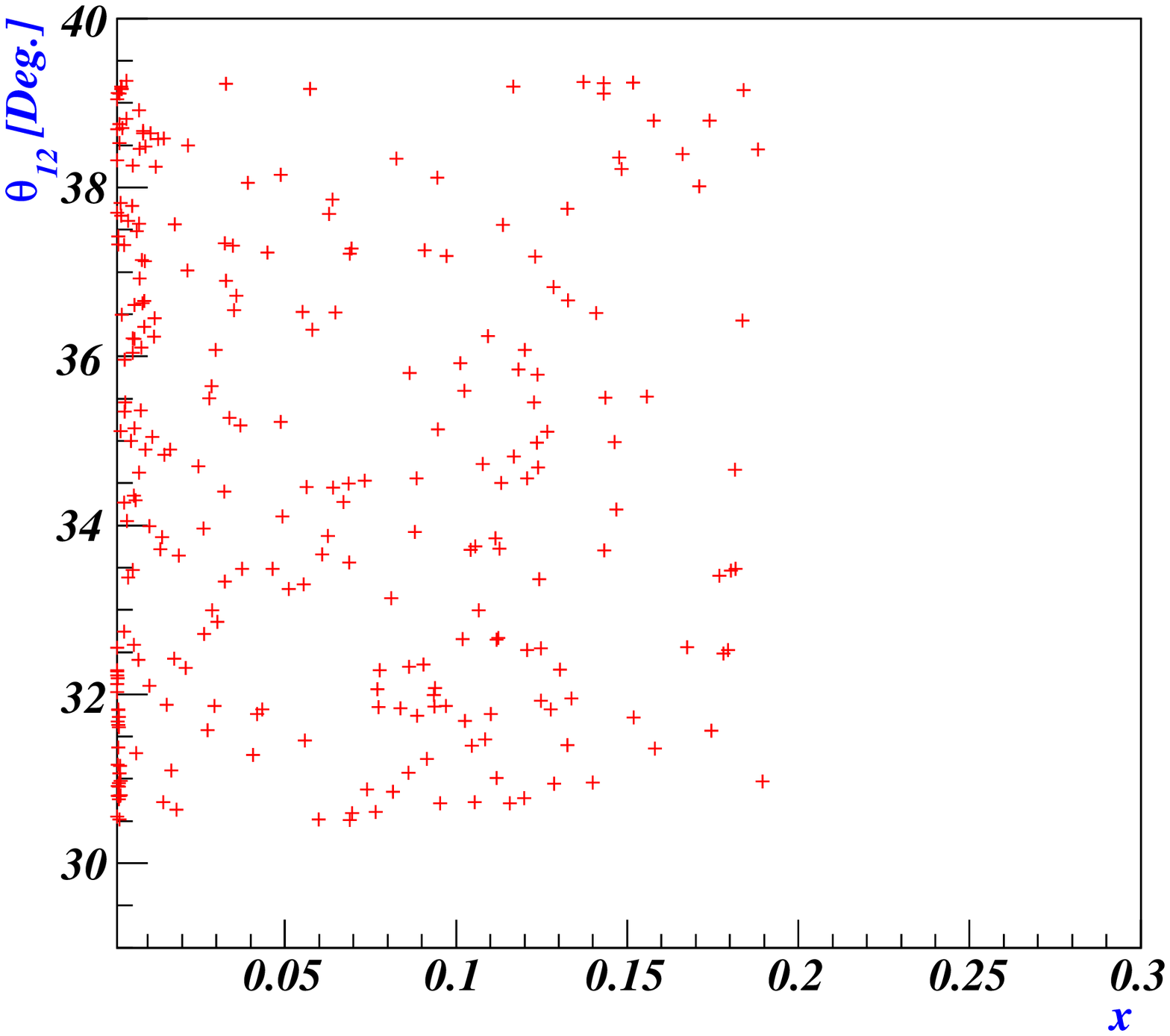,width=6.5cm,angle=0}
\end{minipage}
\hspace*{1.0cm}
\begin{minipage}[t]{6.0cm}
\epsfig{figure=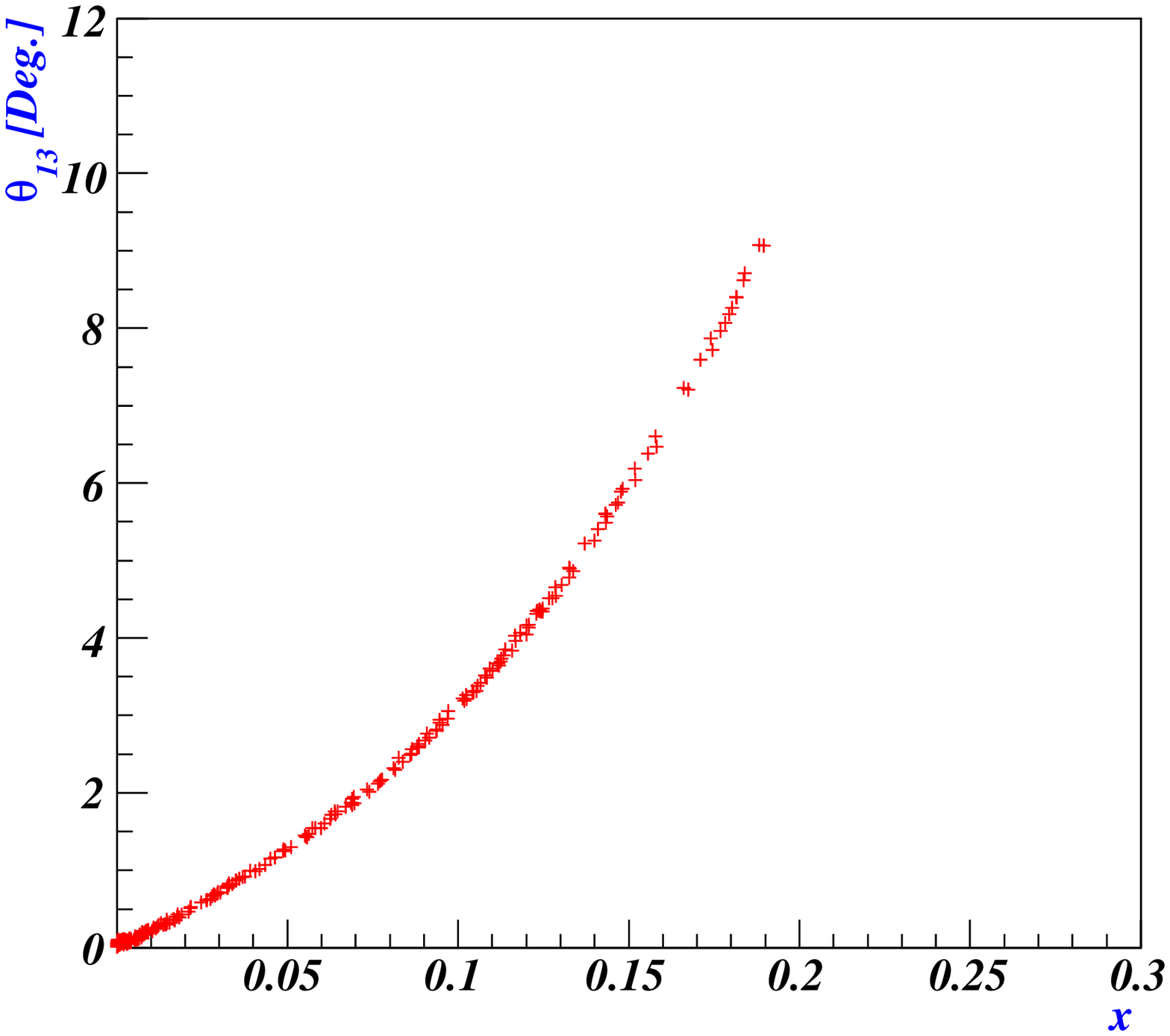,width=6.5cm,angle=0}
\end{minipage}
\caption{\label{Fig5}  The same as Fig.\ref{Fig3}.}
\end{figure}

Fig.~\ref{Fig4} shows how the mixing angles $\theta_{12}, \theta_{13}, \theta_{23}$ and $\delta_{\rm CP}$ in neutrino oscillation depend on the parameter $\phi$, which can be explained in the approximate analysis Eqs.~(\ref{theta23},\ref{theta13},\ref{theta12a}). Especially, Fig.~\ref{Fig4} indicates in the limit of $\phi\rightarrow3\pi/2$ (not in $\phi=3\pi/2$) the value of $\theta_{13}$ can be large.
Fig.~\ref{Fig5} represents that how the mixing angles $\theta_{12}$ and $\theta_{13}$ depend on the parameter $x$, as can be seen in Eqs.~(\ref{theta13}-\ref{theta12}), in which especially the unknown mixing angle $\theta_{13}$ is very sensitive to the parameter $x$.

\section{Leptogenesis}
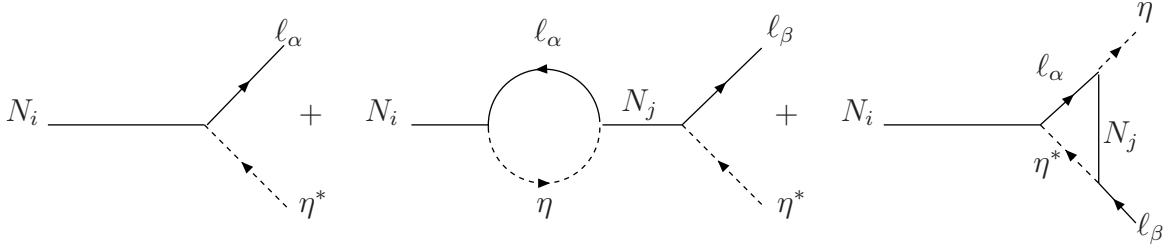
\begin{figure}[hbt]
\begin{picture}(458,100) (30,-5)
\SetWidth{0.5}
\Line(46,40)(105,40)
\DashArrowLine(136,9)(105,40){2}
\Line(183,40)(212,40)
\ArrowArc(233,40)(21.02,3,183)
\Line(255,40)(285,40)
\ArrowLine(285,40)(315,69)
\DashArrowLine(315,10)(285,40){2}
\Line(361,40)(420,40)
\Line(442,59)(442,18)
\ArrowLine(106,40)(135,70)
\Text(30,40)[lb]{\normalsize{$N_{i}$}}
\Text(166,40)[lb]{\normalsize{$N_{i}$}}
\Text(346,40)[lb]{\normalsize{$N_{i}$}}
\Text(263,42)[lb]{\normalsize{$N_{j}$}}
\Text(445,30)[lb]{\normalsize{$N_{j}$}}
\Text(230,70)[lb]{\normalsize{$\ell_{\alpha}$}}
\Text(132,70)[lb]{\normalsize{$\ell_{\alpha}$}}
\Text(318,70)[lb]{\normalsize{$\ell_{\beta}$}}
\Text(141,5)[lb]{\normalsize{$\eta^{\ast}$}}
\Text(322,5)[lb]{\normalsize{$\eta^{\ast}$}}
\ArrowLine(420,40)(442,60)
\DashArrowLine(442,18)(420,40){2}
\DashArrowLine(442,61)(456,75){2}
\ArrowLine(456,3)(442,18)
\Text(141,40)[lb]{\normalsize{$+$}}
\Text(321,40)[lb]{\normalsize{$+$}}
\DashArrowArc(233,39)(21,-180,0){2}
\Text(231,5)[lb]{\normalsize{$\eta$}}
\Text(419,20)[lb]{\normalsize{$\eta^{\ast}$}}
\Text(458,-5)[lb]{\normalsize{$\ell_{\beta}$}}
\Text(420,57)[lb]{\normalsize{$\ell_{\alpha}$}}
\Text(458,79)[lb]{\normalsize{$\eta$}}
\end{picture}
\caption{\label{Fig6} Decay of $N_{i}$ into $\ell_{\alpha L}$ and $\eta$.}
\end{figure}
In addition to the explanation of the smallness of neutrino masses which is generated radiatively through one loop, one of the most popular mechanisms to produce the baryon asymmetry so-called leptogenesis \cite{Fukugita:1986hr} can be explained by introducing singlet heavy Majorana neutrinos. The right-handed heavy Majorana neutrinos decay in the early Universe to a lepton (charged or neutral) and scalar (charged or neutral), thereby  generating a nonzero lepton asymmetry, which in turn gets recycled into a baryon asymmetry through non-perturbative sphaleron processes. We are in the energy scale where $A_{4}$ symmetry is broken but the SM gauge group remains unbroken. So, both the charged and neutral scalars are physical.

The CP asymmetry generated through the interference between tree and one-loop diagrams for the decay of the
heavy Majorana neutrino $N_{i}$ into $\eta$ and $(\nu,\ell_{\alpha})$ is given, for each lepton flavor $\alpha~(=e,\mu,\tau)$, by \cite{lepto2,Flavor}
 \begin{eqnarray}\nonumber
  \varepsilon^{\alpha}_{i} &=& \frac{\Gamma(N_{i}\rightarrow \ell_{\alpha}\eta)
  -\Gamma(N_{i}\rightarrow \overline{\ell}_{\alpha}\eta^{\dag})}{\sum_{\alpha}[\Gamma(N_{i}\rightarrow \ell_{\alpha}\eta)
  +\Gamma(N_{i}\rightarrow\overline{\ell}_{\alpha}\eta^{\dag})]}\\
  &=& \frac{1}{8\pi(\tilde{Y}^{\dag}_{\nu}\tilde{Y}_{\nu})_{ii}}\sum_{j\neq i}{\rm
  Im}\Big\{(\tilde{Y}^{\dag}_{\nu}\tilde{Y}_{\nu})_{ij}(\tilde{Y}_{\nu})^{\ast}_{\alpha i}(\tilde{Y}_{\nu})_{\alpha j}\Big\}g\Big(\frac{M^{2}_{j}}{M^{2}_{i}}\Big),
 \label{cpasym1}
 \end{eqnarray}
where the function $g(x)$ is given by
 \begin{eqnarray}
  g(x)&=& \sqrt{x}\Big[\frac{1}{1-x}+1-(1+x){\rm ln}\frac{1+x}{x}\Big]~.
  \label{decayfunction}
  \end{eqnarray}
Here $i$ denotes a generation index and $\Gamma(N_i \rightarrow \cdot \cdot \cdot)$ is the decay width of the $i$th-generation right-handed neutrino. Note that in our scenario the CP asymmetry is generated by explicitly breaking the tri-bimaximal as in Eq.~(\ref{MR2}) when $x$ is different from zero. Below temperature $T\sim M_{i}\lesssim10^{5}$ GeV, it is known that electron, muon and tau charged lepton Yukawa interactions are much faster than the Hubble expansion parameter rendering the $e$, $\mu$ and $\tau$ Yukawa couplings in equilibrium. Then, the processes which wash out lepton number are flavor dependent and thus the lepton asymmetries for each flavor should be treated separately with different wash-out factors. Once the initial values of $\varepsilon^{\alpha}_{i}$ are fixed, the final result of $\eta_{B}$ or $Y_{B}$ can be obtained by solving a set of flavor-dependent Boltzmann equations including the decay, inverse decay, and scattering processes as well as the nonperturbative sphaleron interaction.

In order to estimate the wash-out effects, we introduce the parameters $K^{\alpha}_{i}$ which are the wash-out factors due to the inverse decay of the Majorana neutrino $N_{i}$ into the lepton flavor $\alpha(=e,\mu,\tau)$ \cite{Abada}. The explicit form of $K^{\alpha}_{i}$ is given by
 \begin{eqnarray}
  K^{\alpha}_{i} =\frac{\Gamma(N_{i}\rightarrow \eta \ell_{\alpha})}{H(M_{i})}
  =K_{i}\frac{(\tilde{Y}^{\ast}_{\nu})_{\alpha i}(\tilde{Y}_{\nu})_{\alpha i}}{(\tilde{Y}^{\dag}_{\nu}\tilde{Y}_{\nu})_{ii}}~,
  \label{K-factor2}
 \end{eqnarray}
 where $\Gamma(N_{i}\rightarrow \eta\ell_{\alpha})$ and $H(M_{i})=(4\pi^{3}g_{\ast}/45)^{\frac{1}{2}}M^{2}_{i}/M_{\rm Pl}$ with the Planck mass $M_{\rm Pl}=1.22\times10^{19}$ GeV and the effective number of degrees of freedom $g_{\ast}\simeq g_{\ast \rm SM}=106.75$ denote the partial decay rate of the process $N_{i}\rightarrow\ell_{\alpha}+\eta$ and the Hubble parameter at temperature $T\simeq M_{i}$, respectively. And the $K$-factors $K_{i}$ associated with $N_{i}$ are given as
 \begin{eqnarray}
  K_{i}\equiv\frac{\Gamma_{N_i}}{H(M_{i})}
  \simeq m_{\ast}\frac{g^{2}_{\nu}}{M_{i}}\delta^{2}_{N_i\eta}~,
  \label{Ki}
 \end{eqnarray}
where $\Gamma_{N_i}$ is a decay width of $N_{i}$ into $\eta$ and $\ell_{\alpha}$ which is defined as $\Gamma_{N_{i}}\equiv\sum_{\alpha}[\Gamma(N_{i}\rightarrow \ell_{\alpha}\eta)
  +\Gamma(N_{i}\rightarrow\overline{\ell}_{\alpha}\eta^{\dag})]=\frac{1}{8\pi}(\tilde{Y}^{\dag}_{\nu}\tilde{Y}_{\nu})_{ii}M_{i}\delta^{2}_{N_i\eta}$ and  $m_{\ast}=\big(\frac{45}{2^{8}\pi^{5}g_{\ast}}\big)^{\frac{1}{2}}M_{\rm Pl}\simeq2.83\times10^{16}$ GeV.
Here the degeneracy between $M^{2}_{i}$ and $\bar{m}^{2}_{\eta}$ is given by~\cite{Gu:2008yk}
 \begin{eqnarray}
  \delta_{N_i\eta}\equiv1-\frac{1}{z_{i}}~,
  \label{degeneracy}
 \end{eqnarray}
where $z_{i}=M^{2}_{i}/\bar{m}^{2}_{\eta}$ and $\delta_{N_i\eta}$ goes to 1 and 0 for $z_{i}\gg1$ and  $z_{i}\rightarrow1$, respectively.

Since the factor $K_{i}$ is dependent on both heavy right-handed neutrino mass $M_{i}$ and Yukawa coupling $g_{\nu}$, which contribution appears as in Eq.~(\ref{mo1}), as well as depends on the degree of degeneracy between $M^{2}_{i}$ and $\bar{m}^{2}_{\eta}$, we can expect that for $M^{2}_{i}\gg\bar{m}^{2}_{\eta}$ the degree of degeneracy $\delta_{N_i\eta}$ is going to be 1, and the produced CP-asymmetries are strongly washed out. In order for this enormously huge wash-out factor to be tolerated, we should consider the case $M^{2}_{i}\approx\bar{m}^{2}_{\eta}$ where the $M_{i}$ is the lightest heavy Majorana neutrino mass, that is $\delta_{N_i\eta}\rightarrow0$. It is clear that, if the value $g^{2}_{\nu}/M_{i}$ is constrained by both low energy neutrino data and LHC signal constraints, the wash-out factors $K^{e}_{i}$ and $K^{\mu\tau}_{i}$ are only dependent on the parameter $\delta_{N_i\eta}$. However, in our scenario we could not provide the explanation of the size of $\delta_{N_i\eta}$.
Here, we note that each CP asymmetry for a single flavor given in Eq.~(\ref{cpasym1}) is weighted differently by the corresponding wash-out parameter given by Eq.~(\ref{K-factor2}), and appears with different weight in the final formula for the baryon asymmetry\cite{Abada};
 \begin{eqnarray}
  \eta_{\emph{B}}&\simeq&
  -2\times10^{-2}\sum_{N_{i}}\Big[\varepsilon^{e}_{i}\tilde{\kappa}\Big(\tiny{\frac{151}{179}}K^{e}_{i}\Big)
  +\varepsilon^{\mu}_{i}\tilde{\kappa}\Big(\tiny\frac{344}{537}K^{\mu}_{i}\Big)
  +\varepsilon^{\tau}_{i}\tilde{\kappa}\Big(\tiny\frac{344}{537}K^{\tau}_{i}\Big)\Big]~,
 \end{eqnarray}
with wash-out factor
 \begin{eqnarray}
  \tilde{\kappa}\simeq\Big(\frac{8.25}{K^{\alpha}_{i}}+\Big(\frac{K^{\alpha}_{i}}{0.2}\Big)^{1.16}\Big)^{-1}~.
 \end{eqnarray}
In our scenario, although $\delta_{N_i\eta}$ does not much affect the results for low energy neutrino observables obtained in sec. III, the predictions of the baryon asymmetry $\eta_{B}$ strongly depends on the quantity $\delta_{N_i\eta}$ due to the size of wash-out parameters. So, we will show the predictions of the baryon asymmetry for the specific values of $\delta_{N_i\eta}$.
And, it seems difficult for resonant leptogenesis to be implemented, because of the constraints of solar mixing angle and the mass-squared differences $\Delta m^{2}_{21}$ and $\Delta m^{2}_{32}$. From the mass-squared differences in Eq.~(\ref{massSqd1}) and Eq.~(\ref{massSqd2}), since for $M_{2}=M_{3}$ (which means $b=1$ and $\omega_{2}=\omega_{3}$) could not give the value of $R\equiv\Delta m^{2}_{21}/|\Delta m^{2}_{32}|\sim{\cal O}(0.01)$, it is not possible for the resonant leptogenesis between $N_{2}$ and $N_{3}$. In the case of degeneracy between $N_{1}$ and $N_{2}$, introducing $\delta_{12}\equiv1-M_{1}/M_{2}$, the solar mixing angle in Eq.~(\ref{theta12a}) indicates $\delta_{12}\gg x\cos\phi$ to satisfy the low energy experimental data, with a large $x$ and a relatively large $\theta_{13}$ at around $\phi=3\pi/2$. However, since its CP-asymmetries are proportional to $\varepsilon^{e}_{1}\sim\sin2\phi,~\varepsilon^{\mu}_{1}\sim\cos\phi$ and $\varepsilon^{\mu}_{1}\sim\cos\phi$, at around $\phi=3\pi/2$ the resonant leptogenesis of the degeneracy between $N_{2}$ and $N_{3}$ could not give a explanation of BAU. In order to explain the possibility of BAU, we will show the two cases corresponding to normal hierarchical and degenerate mass ordering, for example, {\bf Case-I} and {\bf Case-II}, respectively.

{\bf In the case of $M_{3}\gtrapprox\bar{m}_{\eta}$,}
from Eq.~(\ref{Knu}) and Eq.~(\ref{K-factor2}), the wash-out parameters associated with $N_{3}$ and the lepton flavors $\alpha=e,\mu,\tau$ are given as
 \begin{eqnarray}
  K^{e}_{3}= \frac{x^{2}g^{2}_{\nu}m_{\ast}\delta^{2}_{N_3\eta}}{6bM},~~
  K^{\mu}_{3}\simeq\frac{g^{2}_{\nu}m_{\ast}\delta^{2}_{N_3\eta}}{2bM}(1+\frac{2x}{\sqrt{3}}\sin\phi),~~
  K^{\tau}_{3}\simeq\frac{g^{2}_{\nu}m_{\ast}\delta^{2}_{N_3\eta}}{2bM}(1-\frac{2x}{\sqrt{3}}\sin\phi),
  \label{K-N}
  \end{eqnarray}
 in which the common factor $g^{2}_{\nu}m_{\ast}/bM$ is constrained by the overall factor of neutrino mass matrix in Eq.~(\ref{overall}) as $8\pi^{2}m_{\ast}m_{0}/b\bar{m}_{\eta}m_{-}$ with $\bar{m}_{\eta}\simeq M_{3}=bM$, all $K$-factors are evaluated at temperature $T=M_{3}$, and $M_{3}$ is the lightest of the heavy Majorana neutrinos. Note here that wash-out factors associated with $N_{1,2}$ and the lepton flavors $\alpha=e,\mu,\tau$ are enormously huge compared to the factors $K^{e,\mu,\tau}_{3}$, and therefore the generated lepton asymmetries associated with $N_{1,2}$ are strongly washed out due to Eqs.~(\ref{K-factor2},\ref{Ki}).
In a radiative seesaw being plus with $A_{4}$ symmetry, Eq.~(\ref{K-N}) explicitly shows how the Yukawa coupling matrix Eq.~(\ref{Knu}) determined by Eq.~(\ref{mo1}) allows for a heavy Majorana neutrino to decay relatively out of equilibrium, simultaneously protecting the $N_{3}$ lepton number from being washed out, even though large $e,\mu$-and $\tau$-Yukawa couplings to $N_{1,2,3}$ exist. And, the CP asymmetries $\varepsilon^{\alpha}_{3}$ are approximately given\footnote{Due to ${\rm Im}[H_{3j}]=0$ for $\varphi_{1,2}=0$, the relation $\varepsilon^{\mu}_{3}+\varepsilon^{\tau}_{3}=-\varepsilon^{e}_{3}$ is satisfied if they are considered to the order of $x^{3}$.}, for $x\ll1$, by
 \begin{eqnarray}
  \varepsilon^{e}_{3}&=& \frac{bx^{2} g^{2}_{\nu}}{32a\pi}(x\sin\phi+a\sin2\phi)~,\nonumber\\
  \varepsilon^{\mu}_{3} &\simeq& \frac{bx g^{2}_{\nu}}{64a\pi}\{4a\sqrt{3}\cos\phi-x[\sqrt{3}+2a\cos\phi(\sqrt{3}\cos\phi+\sin\phi)]\}~,\nonumber\\
  \varepsilon^{\tau}_{3} &\simeq& \frac{bx g^{2}_{\nu}}{64a\pi}\{-4a\sqrt{3}\cos\phi+x[\sqrt{3}+2a\cos\phi(\sqrt{3}\cos\phi-\sin\phi)]\}~.
  \label{normal1}
 \end{eqnarray}
As can be seen in Eqs.~(\ref{K-N},\ref{normal1}), since the lepton asymmetries in $\mu$ and $\tau$ flavors are equal but opposite in sign to the first order, i.e. $\varepsilon^{\mu}_{3}\approx-\varepsilon^{\tau}_{3}$, satisfying $\varepsilon^{\mu}_{3}+\varepsilon^{\tau}_{3}=-\varepsilon^{e}_{3}$, and the wash-out parameters in $\mu$ and $\tau$ are almost equal $K^{\mu}_{3}\approx K^{\tau}_{3}\gg K^{e}_{3}$, the effects of wash-out factor related with $N_{3}$ can play a crucial role in a successful leptogenesis according to the size of $\delta_{N_3\eta}$.
\begin{figure}[t]
\hspace*{-2cm}
\begin{minipage}[t]{6.0cm}
\epsfig{figure=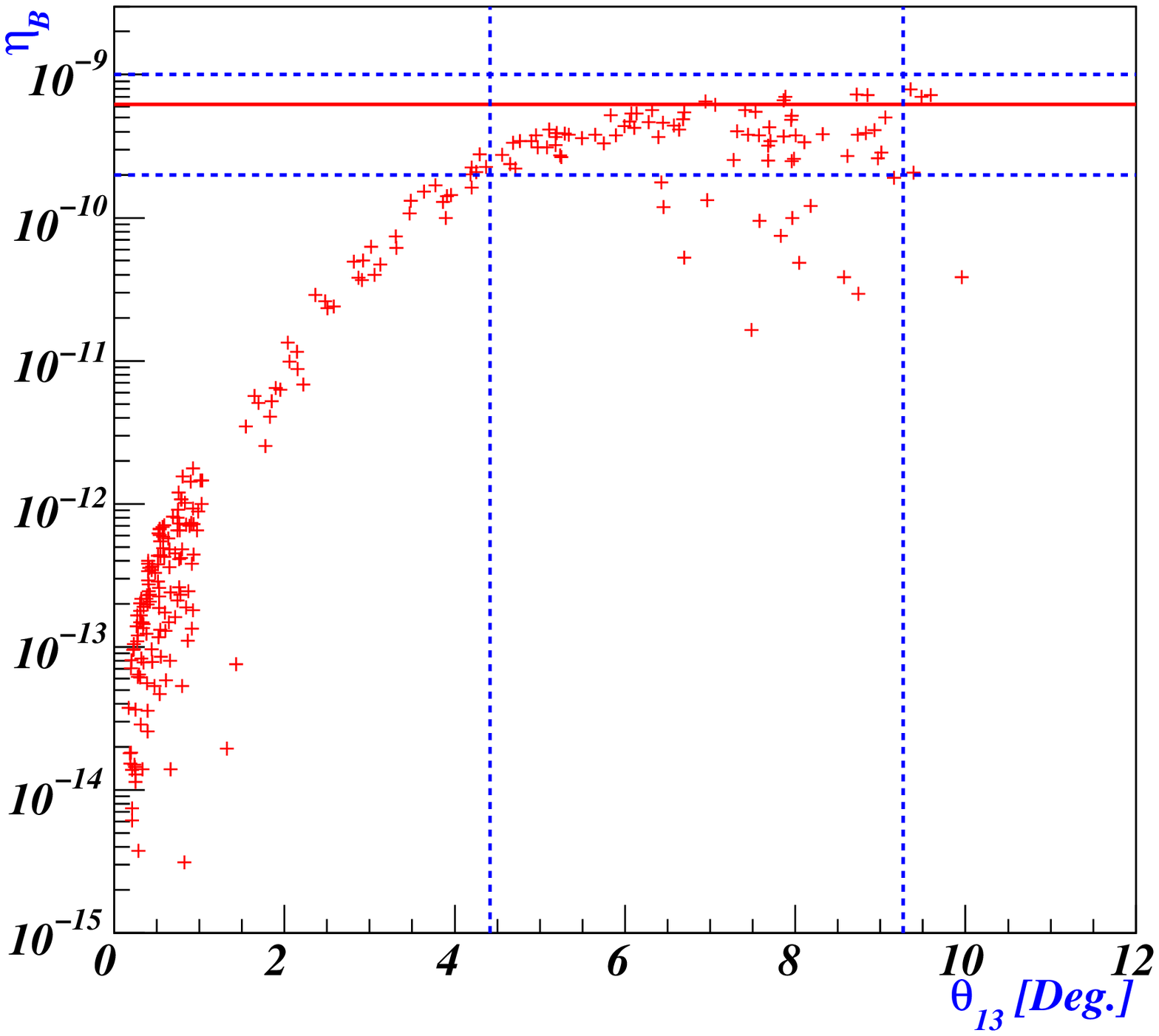,width=6.5cm,angle=0}
\end{minipage}
\hspace*{1.0cm}
\begin{minipage}[t]{6.0cm}
\epsfig{figure=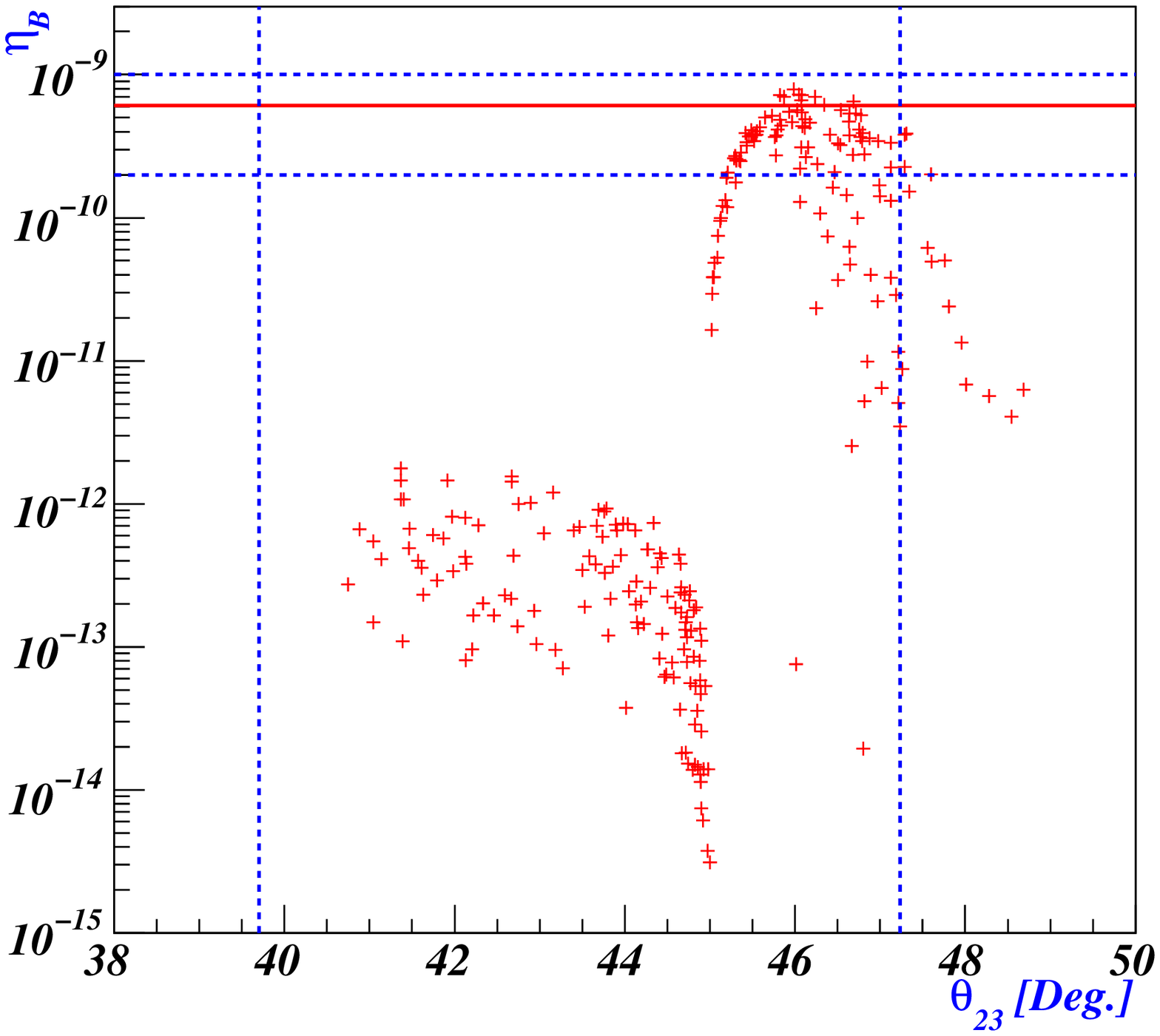,width=6.5cm,angle=0}
\end{minipage}
\caption{\label{Fig7-1} Figures show the predictions of $\eta_B$ for $M_{3}\simeq1{\rm TeV}$ and $\delta_{N_{3}\eta}=10^{-5}$. Left-figure shows $\eta_{B}$ as a function of $\theta_{13}$. Right-figure shows $\eta_{B}$ as a function of $\theta_{23}$. The horizontal
dotted lines in both figures correspond to the phenomenologically acceptable current measurement, and the horizontal thick line represents the best-fit value $\eta_{B}=6.2\times10^{-10}$ of current measurement from WMAP \cite{cmb}. And the vertical dotted lines represent the experimental bounds in $1\sigma$ of the mixing angles $\theta_{13}$ and $\theta_{23}$ in neutrino oscillations.}
\end{figure}
In strong wash-out regime $K^{\alpha}_{3}>1~(\alpha=e,\mu,\tau)$, given the initial thermal abundance of $N_{3}$ and the condition for $K^{\alpha}_{3}$, the resulting baryon-to-photon ratio $\eta_{B}$ including lepton flavor effects can be approximately given as
 \begin{eqnarray}
  \eta_{\emph{B}}&\simeq& 4\times10^{-3}\varepsilon^{e}_{3}\Big(\frac{1}{K^{e}_{3}}\Big)^{1.16}~.
  \label{3st}
 \end{eqnarray}
which indicates electron-flavor effect plays a crucial role in reproducing BAU.
If we take, for example, $\delta_{N_{3}\eta}=10^{-5}$ the magnitude of washout factors are given as
 \begin{eqnarray}
  K^{e}_{3}=0.0006-1.7~,~~~~~K^{\mu}_{3}=55-101~,~~~~~K^{\tau}_{3}=61-124~,
 \end{eqnarray}
and Fig.~\ref{Fig7-1} shows the behavior of $\eta_{B}$ as functions of $\theta_{13}$ and $\theta_{23}$, representing a successful leptogenesis in the range of $1\sigma$ neutrino oscillation data.

\begin{figure}[t]
\hspace*{-2cm}
\begin{minipage}[t]{6.0cm}
\epsfig{figure=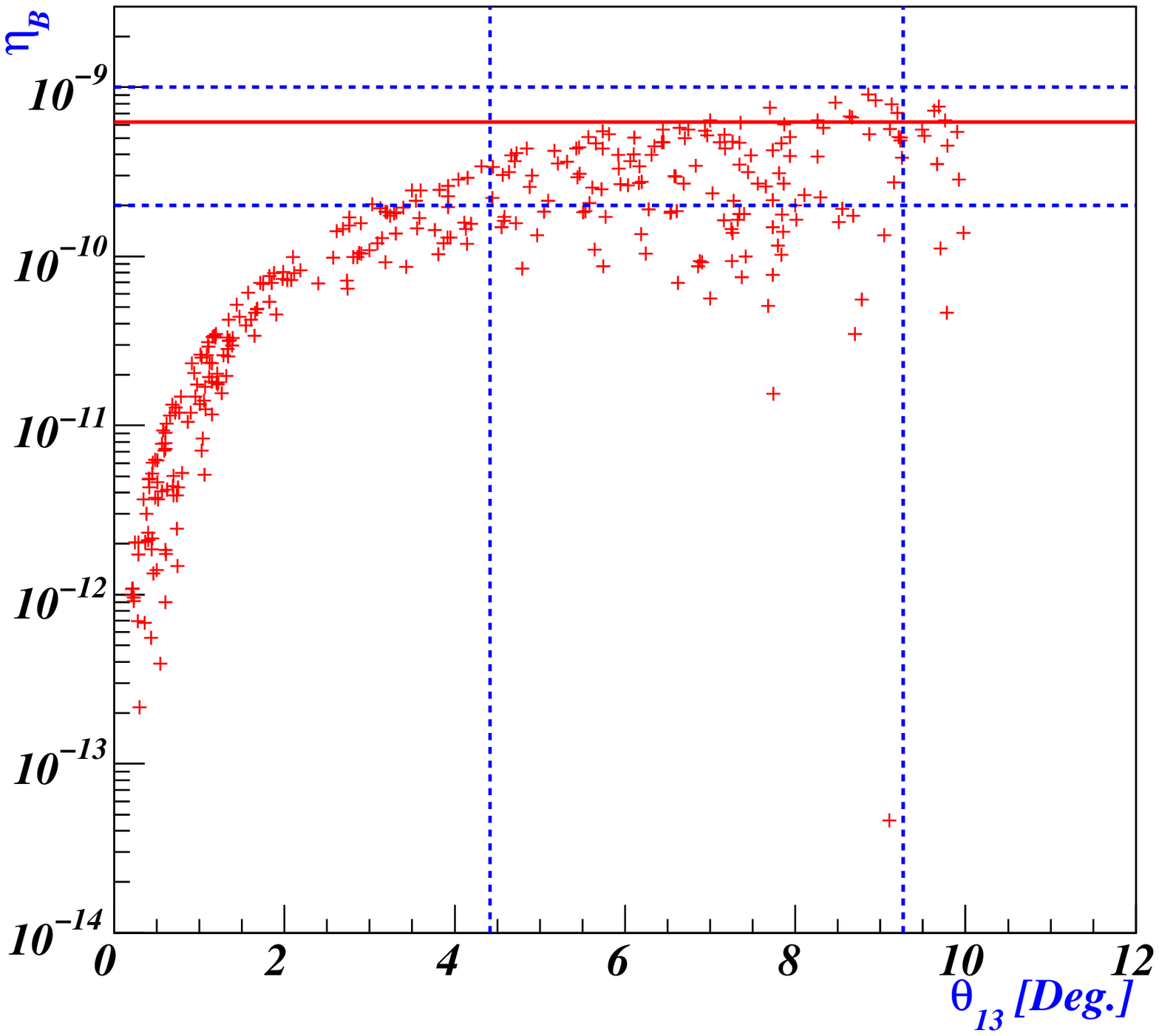,width=6.5cm,angle=0}
\end{minipage}
\hspace*{1.0cm}
\begin{minipage}[t]{6.0cm}
\epsfig{figure=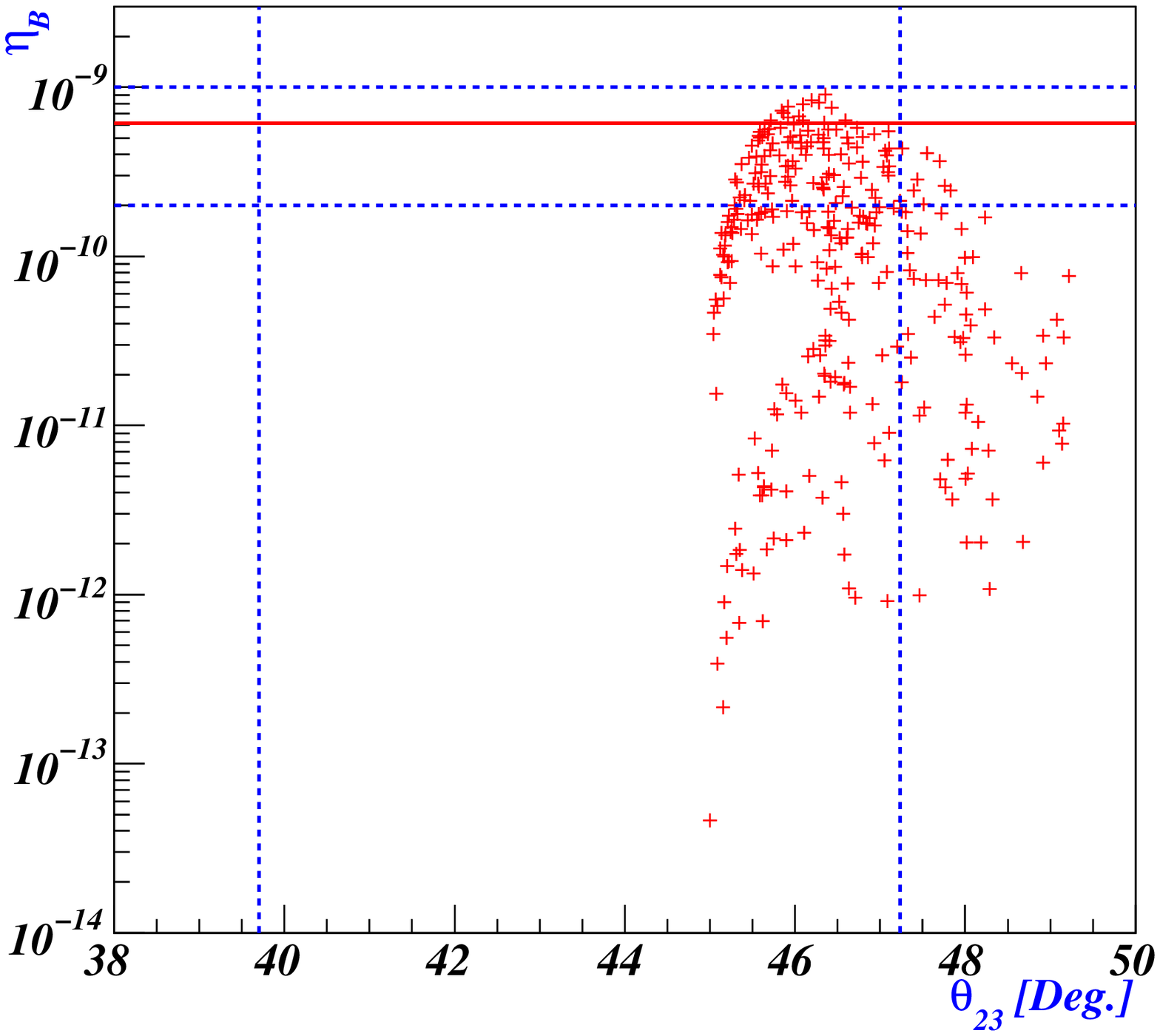,width=6.5cm,angle=0}
\end{minipage}
\caption{\label{Fig7-2} The same sa Fig.~\ref{Fig7-1} except for $M_{3}\simeq1{\rm TeV}$ and $\delta_{N_{3}\eta}=10^{-6}$.}
\end{figure}

In weak wash-out regime $K^{e}_{3}\ll K^{\mu\tau}_{3}<1$, assumed that $N_{3}$ is not initially present in the plasma, but they are generated by the inverse decays and scatterings, the resulting baryon-to-photon ratio $\eta_{B}$ including lepton flavor effects can be simply given as
 \begin{eqnarray}
  \eta_{\emph{B}}&\simeq& 2\times10^{-3}x\sin\phi K_{3}K^{\mu}_{3}(\varepsilon^{\mu}_{3}-\varepsilon^{\tau}_{3})~,
  \label{3wk}
 \end{eqnarray}
 where in this approximation $\varepsilon^{\mu}_{3}+\varepsilon^{\tau}_{3}=-\varepsilon^{e}_{3}$ is used, and $K_{3}=K^{e}_{3}+K^{\mu}_{3}+K^{\tau}_{3}$. This case indicates muon- and tau-flavor effects play a crucial role in reproducing BAU. For example, for $\delta_{N_{3}\eta}=10^{-6}$ the magnitude of washout factors are given as
 \begin{eqnarray}
  K^{e}_{3}=0.00001-0.017~,~~~~~K^{\mu}_{3}=0.5-1.1~,~~~~~K^{\tau}_{3}=0.6-1.3~,
 \end{eqnarray}
and Fig.~\ref{Fig7-2} shows the behavior of $\eta_{B}$ as functions of $\theta_{13}$ and $\theta_{23}$, representing a successful leptogenesis in the range of $1\sigma$ neutrino oscillation data.

{\bf In the case of $M_{2}\gtrapprox\bar{m}_{\eta}$,}
for $M_{1}>M_{3}>M_{2}\gtrsim\bar{m}_{\eta}$ which corresponds to degenerate normal mass ordering of light neutrinos, from Eq.~(\ref{Knu}) and Eq.~(\ref{K-factor2}), the wash-out parameters associated with $N_{2}$ and the lepton flavors $\alpha=e,\mu,\tau$ are given as
 \begin{eqnarray}
  &&K^{e}_{2} \simeq \frac{g^{2}_{\nu}m_{\ast}}{3M}(1+2x\cos\phi)\delta^{2}_{N_2\eta}~,\nonumber\\
  &&K^{\mu}_{2} \simeq \frac{g^{2}_{\nu}m_{\ast}}{3M}(1-x\cos\phi-x\sqrt{3}\sin\phi)\delta^{2}_{N_2\eta}~,\nonumber\\
  &&K^{\tau}_{2} \simeq \frac{g^{2}_{\nu}m_{\ast}}{3M}(1-x\cos\phi+x\sqrt{3}\sin\phi)\delta^{2}_{N_2\eta}~,
  \label{K-N2}
  \end{eqnarray}
where the common factor $\frac{g^{2}_{\nu}m_{\ast}}{M}\simeq8\pi^{2}m_{\ast}m_{0}/(\bar{m}_{\eta}m_{-})$, all $K$-factors are evaluated at temperature $T=M_{2}$, and $M_{2}$ is the lightest of the heavy Majorana neutrinos.
And the corresponding CP asymmetries $\varepsilon^{\alpha}_{2}$ are approximately given, for $x\ll1$, by
 \begin{eqnarray}
  \varepsilon^{e}_{2}&=& \frac{(b-a)x^{2} g^{2}_{\nu}}{32ab\pi}\sin2\phi~,\nonumber\\
  \varepsilon^{\mu}_{2} &\simeq& \frac{x g^{2}_{\nu}}{32ab\pi}\{-2a\sqrt{3}+x(a-b)(\sqrt{3}\cos\phi+\sin\phi)\}\cos\phi~,\nonumber\\
  \varepsilon^{\tau}_{2} &\simeq& \frac{x g^{2}_{\nu}}{32ab\pi}\{2a\sqrt{3}-x(a-b)(\sqrt{3}\cos\phi-\sin\phi)\}\cos\phi~,
  \label{normal2}
 \end{eqnarray}
in which $\varepsilon^{\mu}_{2}+\varepsilon^{\tau}_{2}=-\varepsilon^{e}_{2}$ is satisfied due to ${\rm Im}[H_{2j}]=0$ for $\varphi_{1,2}=0$. From Eq.~(\ref{K-N2}) we see that all $K$-factors are almost equal, so this case can be classified as $K^{\alpha}_{2}\geq1$ and $K^{\alpha}_{2}<1$.
In strong wash-out regime $K^{\alpha}_{2}\geq1~(\alpha=e,\mu,\tau)$, given the initial thermal abundance of $N_{2}$ and the condition for $K^{\alpha}_{2}$, the resulting baryon-to-photon ratio $\eta_{B}$ approximately given as
 \begin{eqnarray}
  \eta_{\emph{B}}\simeq10^{-2}(\varepsilon^{\tau}_{2}-\varepsilon^{\mu}_{2})\frac{x\sin\phi}{(K^{\mu}_{2})^{1.16}}~,
 \label{etaB2a}
 \end{eqnarray}
where $x\sin\phi$ is from the common factor in $K^{\mu,\tau}_{2}$. In weak wash-out regime $K^{\alpha}_{2}<1~(\alpha=e,\mu,\tau)$, the resulting baryon-to-photon ratio $\eta_{B}$ can be simply given as
 \begin{eqnarray}
  \eta_{\emph{B}}\simeq-3.2\times10^{-3}(\varepsilon^{\tau}_{2}-\varepsilon^{\mu}_{2})K_{2}K^{\mu}_{2}x\sin\phi~,
 \label{etaB2b}
 \end{eqnarray}
where $x\sin\phi$ comes from the common factor in $K^{\mu,\tau}_{2}$, $K^{\mu}_{2}\simeq K^{\tau}_{2}$ and $K_{2}=K^{e}_{2}+K^{\mu}_{2}+K^{\tau}_{2}$.

\begin{figure}[t]
\hspace*{-2cm}
\begin{minipage}[t]{6.0cm}
\epsfig{figure=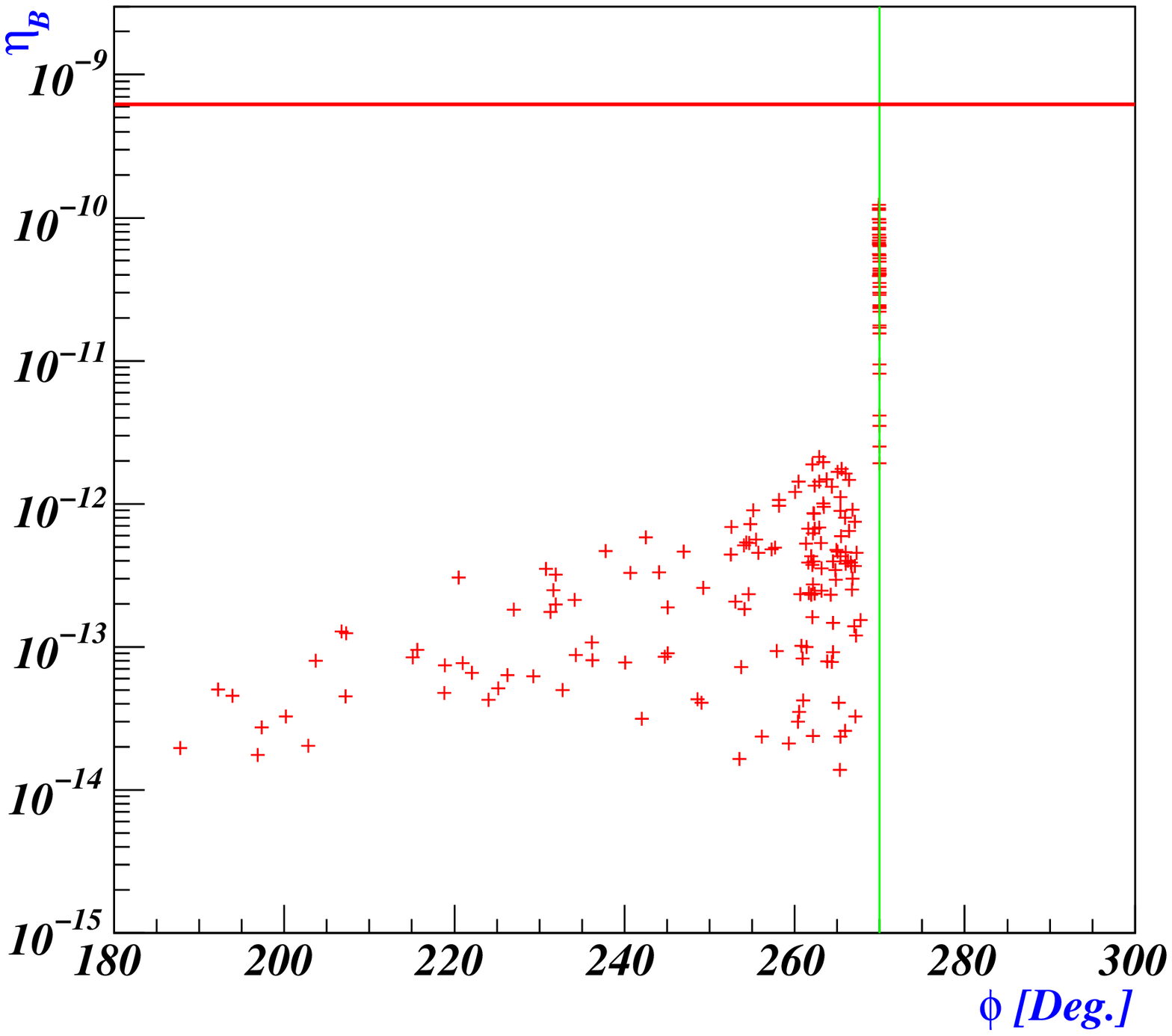,width=6.5cm,angle=0}
\end{minipage}
\hspace*{1.0cm}
\begin{minipage}[t]{6.0cm}
\epsfig{figure=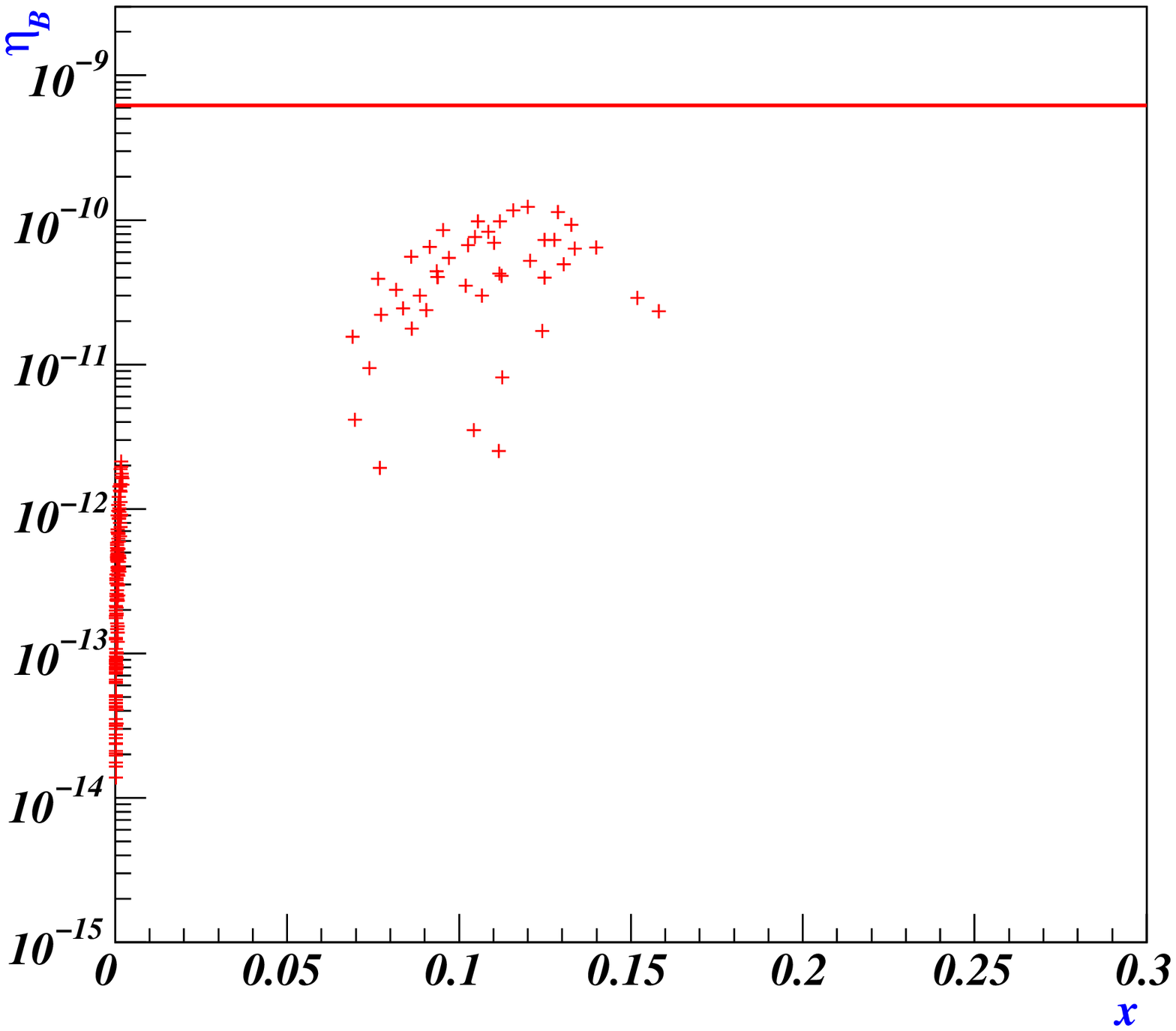,width=6.5cm,angle=0}
\end{minipage}
\caption{\label{Fig8} The same as Fig.\ref{Fig5} except for $M_{2}=1{\rm TeV}$ and $\delta_{N_{2}\eta}=1.5\times10^{-6}$.}
\end{figure}

If we take, for example, $\delta_{N_{3}\eta}=1.5\times10^{-6}$, the magnitude of washout factors are given as
 \begin{eqnarray}
  K^{e}_{3}=3-3.7~,~~~~~K^{\mu}_{3}=3-4.9~,~~~~~K^{\tau}_{3}=2.2-3.7~.
 \end{eqnarray}
and Eq.~(\ref{etaB2a}) can be simplified as
 \begin{eqnarray}
  \eta_{\emph{B}}\simeq4\times10^{-6}x^{2}\sin2\phi~,
 \end{eqnarray}
where $g_{\nu}\simeq0.4,~b\simeq2.7$ and $(K^{\mu}_{2})^{-1.16}\simeq0.2$. Fig.~\ref{Fig8} shows the behavior of $\eta_{B}$ as functions of $\phi$ and $x$, representing only in the limit of $\phi=3\pi/2$ the value of $x$ can become order of $\mathcal{O}(0.1)$. However, Fig.~(\ref{Fig8}) shows that $\eta_{B}$ is not enough for BAU to be satisfied.

\section{Conclusion}

Models based on $A_{4}$ flavor symmetry seems to be extremely attractive because of their predictions of TBM in leading order and naturalness. However, the recent analysis based on global fits of the available data gives us hints for $\theta_{13}>0$ at 1$\sigma$~\cite{bari}, a relatively large $|U_{e3}|=0.126$(best-fit value), even which is not yet a significant indication for a non-zero $U_{e3}$. And this non-zero $U_{e3}$ implies that the TBM $\sin^{2}\theta_{12}=1/3$ is disfavored in $1\sigma$ experimental results due to the upper bound $\sin^{2}\theta_{12}=0.331$ less than $1/3$. Moreover, those models realized on type-I seesaw lead to vanishing leptonic CP-asymmetries responsible for leptogenesis due to the combination $Y^{\dag}_{\nu}Y_{\nu}$ being proportional to the unit matrix. Therefore, in order for non zero $U_{e3}$ and non-vanishing $Y^{\dag}_{\nu}Y_{\nu}$ to be generated with its scale for leptogenesis being predicted at $\sim10^{13-14}$ GeV, higher dimensional operators should be considered \cite{Jenkins:2008rb, Lin:2009bw}.

We have considered an effective theory with an $A_{4}\times Z_{2}\times Z_{4}$ symmetry and investigated the possibility of a linking  TeV-leptogenesis with a relatively large reactor angle through $A_{4}$ symmetry breaking which is at a scale much higher than electroweak scale under the framework of a radiative seesaw. We showed that the non-zero $U_{e3}$ can be generated by adding one five-dimensional effective operator with cut-off scale $\Lambda$, which is responsible for the deviation of the exact TBM, to explain leptogenesis.
We assumed $\Lambda$ to be the CP violation scale which is expected to be much higher than electroweak and $A_4\times Z_4$ symmetry breaking scales. At the very high energy scale $>\Lambda$, the leptonic sector leads to the exact TBM. By introducing a Yukawa interaction between right-handed neutrinos and a $A_4$ triplet scalar field, we obtained non-degenerate heavy Majorana neutrino mass spectrums. In the framework of radiative seesaw that neutrino masses are produced at one-loop level, we concentrated on the effects of CP phase appeared in $Y_{\nu}$, even CP phases coming from both $Y_{\nu}$ and $M_{R}$ participate in the forming of low energy observables and flavored leptogenesis. And we scanned all the parameter space by considering the experimental bounds of low-energy neutrino oscillation data. We analyzed possible spectrums of light neutrinos and their flavor mixing angles corresponding to heavy Majorana neutrino mass ordering, and we found only normal hierarchical and quasi-degenerate spectrums of light neutrino are preferred in our model. The extent of $\theta_{13}$ and Dirac CP phase $\delta$ are also investigated where the size of $\theta_{13}$ is sensitive to ratio $\upsilon_{\psi}/\Lambda$. In particular, in order to show a successful leptogenesis as well as a linking leptogenesis with low energy observables, we have considered only in the case of a non-vanishing CP phase appeared in $Y_{\nu}$, and studied the viability of thermal leptogenesis at TeV scale. Furthermore it turned out that resonant enhancement to lower down the scale of leptogenesis does not work in our scenario. Instead, we considered the phase space suppression method where we used the parameter which represents the degeneracy between lightest $Z_2$-odd scalar and lightest right-handed Majorana neutrino to modulate the wash-out effects in the decaying processes of heavy Majorana neutrinos, and it showed that only hierarchical mass ordering of heavy Majorana neutrino corresponding to normal hierarchical light neutrino mass spectrum could give a successful leptogenesis with a relatively large $U_{e3}$ in our model.
\acknowledgments{
YHA is supported by the National Science Council of R.O.C. under Grants No:
 NSC-97-2112-M-001-004-MY3.}

\newpage
\appendix

\section{Higgs Potential and vacuum alignment}
Since it is not trivial to ensure that the different vacuum alignments of $\langle\Phi^{0}\rangle=(\upsilon,\upsilon,\upsilon)$, $\langle\vartheta\rangle=\upsilon_{\vartheta}$ and $\langle\chi\rangle,\langle\psi\rangle$ in Eq.~(\ref{subgroup}) are preserved, or at least approximately preserved, we shall briefly discuss these vacuum alignments.
In order for the different vacuum alignments of $\langle\Phi\rangle,\langle\chi\rangle,\langle\vartheta\rangle$ and $\langle\psi\rangle$ to be ensured, let us consider the most general renormalizable scalar potential of $\Phi,\chi,\vartheta,\psi$ and $\eta$ invariant under $SU(2)\times U(1)\times A_{4}$ with the $Z_{2}\times Z_{4}$ symmetry is given as
\newcommand{\3}{\mbox{${\bf \underline{3}}$}}
\newcommand{\s}{\mbox{${\bf \underline{1}}$}}
\newcommand{\spr}{\mbox{${\bf \underline{1}'}$}}
\newcommand{\sppr}{\mbox{${\bf {\underline{1}''}}$}}
\begin{eqnarray}
V &=& \mu^2_\Phi B_{1} +\lambda^\Phi_1 Q^{\Phi}_{1} + \lambda^\Phi_2 Q^{\Phi}_{2}+\lambda^\Phi_3 Q^{\Phi}_{3} + \lambda^\Phi_4 Q^{\Phi}_{4}+ i \lambda^\Phi_5 Q^{\Phi}_{5}\nonumber\\
&+& \mu^2_\chi B_{2}  + \lambda^\chi_1 Q^{\chi}_{1} + \lambda^\chi_2 Q^{\chi}_{2}+ \lambda^\chi_3 Q^{\chi}_{3}+
\mu^2_\psi B_{3} +\delta^\psi T_{1} + \lambda^\psi_1 Q^{\psi}_{1} + \lambda^\psi_2 Q^{\psi}_{2} + \lambda^\psi_3 Q^{\psi}_{3}\nonumber\\
&+& \mu^2_\vartheta B_{4} + \lambda^\vartheta Q^{\vartheta}_{1}
+\lambda^{\Phi\chi}_1 Q^{\Phi\chi}_{1}+\lambda^{\Phi\chi}_2 Q^{\Phi\chi}_{2}+\lambda^{\Phi\chi*}_2
Q^{\Phi\chi}_{3}+\lambda^{\Phi\chi}_3 Q^{\Phi\chi}_{4} +i\lambda^{\Phi\chi}_4 Q^{\Phi\chi}_{5}\nonumber\\
 &+& \delta^{\Phi\psi}_{s}T_{2} + i \delta^{\Phi\psi}_a T_{3}+\lambda^{\Phi\psi}_1 Q^{\Phi\psi}_{1}+\lambda^{\Phi\psi}_2 Q^{\Phi\psi}_{2}+\lambda^{\Phi\psi*}_2 Q^{\Phi\psi}_{3}+\lambda^{\Phi\psi}_3 Q^{\Phi\psi}_{4} +i\lambda^{\Phi\psi}_4 Q^{\Phi\psi}_{5}\nonumber\\
 &+&  \lambda^{\Phi\vartheta}Q^{\Phi\vartheta}_{1} + \delta^{\psi\chi}T_{4}+\lambda^{\psi\chi}_{1} Q^{\psi\chi}_{1}+\lambda^{\psi\chi}_{2}Q^{\psi\chi}_{2}+\lambda^{\psi\chi\ast}_{2}Q^{\psi\chi}_{3}+\lambda^{\psi\chi}_{3}Q^{\psi\chi}_{4}\nonumber\\
&+& \lambda^{\psi\vartheta}Q^{\psi\vartheta}_{1}+ \delta^{\chi\vartheta\psi}T_{5}+\mu^2_\eta B_{5}  + \lambda^\eta Q^{\eta}_{1}+ \lambda^{\Phi\eta}_1 Q^{\Phi\eta}_{1} + \lambda^{\Phi\eta}_2 Q^{\Phi\eta}_{2} + \lambda^{\Phi\eta}_3 Q^{\Phi\eta}_{3} +\lambda^{\Phi\eta*}_3 Q^{\Phi\eta}_{4}\nonumber\\
&+& \lambda^{\eta\chi}Q^{\eta\chi}_{1}+\lambda^{\eta\psi}Q^{\eta\psi}_{1}+\lambda^{\eta\vartheta}Q^{\eta\vartheta}_{1}~,
\end{eqnarray}
where the bilinears $B$, trilinears $T$ and quartic terms $Q$ are given as
 \begin{eqnarray}
  B_{1} &=&(\Phi^\dagger \Phi)_{\s},\quad~~~~~
  B_{2} = (\chi \chi)_{\s},\quad~~~~~~~
  B_{3} = (\psi \psi)_{\s},\quad~~~~~~
  B_{4} = \vartheta \vartheta,\nonumber\\
  B_{5} &=& \eta^\dagger \eta,\quad~~~~~~~~~~~
  T_{1} = (\psi\psi\psi)_{\s},\quad~~~~~
  T_{2} = (\Phi^\dagger \Phi)_{\3s}\psi,\quad~~
  T_{3} = (\Phi^\dagger \Phi)_{\3a}\psi,\nonumber\\
  T_{4} &=& (\psi\chi\chi)_{\s},\quad~~~~~
  T_{5} = (\chi\psi)_{\s}\vartheta,
 \end{eqnarray}
 \begin{eqnarray}
  Q^{\Phi}_{1} &=& (\Phi^\dagger \Phi)_{\s}(\Phi^\dagger \Phi)_{\s},\quad~~~~~
  Q^{\Phi}_{2} = (\Phi^\dagger \Phi)_{\s'}(\Phi^\dagger \Phi)_{\s''},\quad~~~
  Q^{\Phi}_{3} = (\Phi^\dagger \Phi)_{\3s}(\Phi^\dagger\Phi)_{\3s},\nonumber\\
  Q^{\Phi}_{4} &=& (\Phi^\dagger\Phi)_{\3a}(\Phi^\dagger \Phi)_{\3a},\quad~~~
  Q^{\Phi}_{5} = (\Phi^\dagger \Phi)_{\3s}(\Phi^\dagger\Phi)_{\3a},\quad~~~
  Q^{\chi}_{1} = (\chi\chi)_{\s}(\chi\chi)_{\s},\nonumber\\
  Q^{\chi}_{2} &=& (\chi\chi)_{\s'}(\chi\chi)_{\s''},\quad~~~~~~~
  Q^{\chi}_{3} = (\chi\chi)_{\3}(\chi\chi)_{\3},\quad~~~~~~~~
  Q^{\psi}_{1} = (\psi\psi)_{\s}(\psi\psi)_{\s},\nonumber\\
  Q^{\psi}_{2} &=& (\psi\psi)_{\s'}(\psi\psi)_{\s''},\quad~~~~~~
  Q^{\psi}_{3} = (\psi\psi)_{\3}(\psi\psi)_{\3},\quad~~~~~~~~
  Q^{\vartheta}_{1} = (\vartheta\vartheta)^2,\nonumber\\
  Q^{\Phi\chi}_{1} &=& (\Phi^\dagger\Phi)_{\s}(\chi\chi)_{\s},\quad~~~~~~~
  Q^{\Phi\chi}_{2} = (\Phi^\dagger\Phi)_{\s'}(\chi\chi)_{\s''},\quad~~~~
  Q^{\Phi\chi}_{3} = (\Phi^\dagger\Phi)_{\s''}(\chi\chi)_{\s'},\nonumber\\
  Q^{\Phi\chi}_{4} &=& (\Phi^\dagger\Phi)_{\3s}(\chi\chi)_{\3},\quad~~~~~~
  Q^{\Phi\chi}_{5} = (\Phi^\dagger\Phi)_{\3a}(\chi\chi)_{\3},
 \end{eqnarray}
 \begin{eqnarray}
  Q^{\Phi\psi}_{1} &=& (\Phi^\dagger\Phi)_{\s}(\psi\psi)_{\s},\quad~~
  Q^{\Phi\psi}_{2} = (\Phi^\dagger\Phi)_{\s'}(\psi\psi)_{\s''},\quad~~
  Q^{\Phi\psi}_{3} = (\Phi^\dagger\Phi)_{\s''}(\psi\psi)_{\s'},\nonumber\\
  Q^{\Phi\psi}_{4} &=& (\Phi^\dagger\Phi)_{\3s}(\psi\psi)_{\3},\quad~
  Q^{\Phi\psi}_{5} = (\Phi^\dagger\Phi)_{\3a}(\psi\psi)_{\3},\quad~~~
  Q^{\Phi\vartheta}_{1} = (\Phi^\dagger\Phi)_{\s}(\vartheta\vartheta),\nonumber\\
  Q^{\psi\chi}_{1} &=& (\psi\psi)_{\s}(\chi\chi)_{\s},\quad~~~~
  Q^{\psi\chi}_{2} = (\psi\psi)_{\s'}(\chi\chi)_{\s''},\quad~~~~
  Q^{\psi\chi}_{3} = (\psi\psi)_{\s''}(\chi\chi)_{\s'},\nonumber\\
  Q^{\psi\chi}_{4} &=& (\psi\psi)_{\3}(\chi\chi)_{\3},\quad~~~~
  Q^{\psi\vartheta}_{1} = (\psi\psi)_{\s}(\vartheta\vartheta),\quad~~~~~~~~
  Q^{\eta}_{1} =(\eta^\dagger\eta)^2,\nonumber\\
  Q^{\Phi\eta}_{1} &=&(\Phi^\dagger\Phi)_{\s}(\eta^\dagger \eta),\quad~~~~
  Q^{\Phi\eta}_{2} =(\Phi^\dagger\eta)(\eta^\dagger \Phi),\quad~~~~~~~~
  Q^{\Phi\eta}_{3} =(\Phi^\dagger\eta)(\Phi^\dagger \eta),\nonumber\\
  Q^{\Phi\eta}_{4} &=&(\eta^\dagger \Phi)(\eta^\dagger \Phi),\quad~~~~~~
  Q^{\eta\chi}_{1} = (\eta^\dagger \eta)(\chi\chi)_{\s},\quad~~~~~~~~
  Q^{\eta\psi}_{1} = (\eta^\dagger\eta)(\psi\psi)_{\s},\nonumber\\
  Q^{\eta\vartheta}_{1} &=&(\eta^\dagger\eta)(\vartheta\vartheta)~.
 \end{eqnarray}
The field configurations in our scenario are assumed to be as
\begin{eqnarray}
\langle \Phi \rangle = (\upsilon,\upsilon,\upsilon)~,\quad \langle\chi\rangle = (\upsilon_{\chi},0,0)~, \quad\langle\Psi\rangle = (0,\upsilon_{\psi},0)~, \quad\langle\vartheta\rangle = \upsilon_{\vartheta}~, \quad \langle \eta \rangle = 0~.
\end{eqnarray}
The presence of interactions including the terms like $Q^{\Phi\chi}_{1,...,4}$, $Q^{\Phi\psi}_{1,...,4}$, $\delta^{\Phi\psi}_{s,a}$ and $\delta^{\chi\vartheta\psi}$ supply a large number of independent equations of extremum conditions than there are unknown VEVs ($\upsilon,\upsilon_{\chi},\upsilon_{\psi}$ and $\upsilon_{\vartheta}$), which means that unnatural fine-tuning conditions have to be enforced on the Higgs potential parameters.
Using the extremum conditions, we obtain
 \begin{eqnarray}
  0=\frac{\partial V}{\partial\varphi_{1}} &=& 2\mu^2_{\Phi}v + 4(3\lambda_1^{\Phi} + 4\lambda_3^{\Phi})v^3 + 2(\lambda_1^{\Phi\chi} + \lambda_2^{\Phi\chi} + \lambda_2^{\Phi\chi*})vv_{\chi}^2 \nonumber \\
  && + 2(\lambda_1^{\Phi\psi} + \omega^2\lambda_2^{\Phi\psi}
  + \omega\lambda_2^{\Phi\psi*})vv_{\psi}^2
 + (\delta_s^{\Phi\psi} + i\delta_a^{\Phi\psi})vv_{\psi} + 2\lambda^{\Phi\vartheta}vv^2_{\vartheta} ~,~\nonumber\\
  0=\frac{\partial V}{\partial\varphi_{2}} &=& 2\mu^2_{\Phi}v +  4(3\lambda_1^{\Phi} + 4\lambda_3^{\Phi})v^3 + 2(\lambda_1^{\Phi\chi} + \omega\lambda_2^{\Phi\chi} + \omega^{2}\lambda_2^{\Phi\chi*})vv_{\chi}^2 \nonumber \\
  && + 2(\lambda_1^{\Phi\psi} + \lambda_2^{\Phi\psi} + \lambda_2^{\Phi\psi*})vv_{\psi}^2
 + 2\lambda^{\Phi\vartheta}vv^2_{\vartheta} ~,~\nonumber\\
  0=\frac{\partial V}{\partial\varphi_{3}} &=& 2\mu^2_{\Phi}v + 4(3\lambda_1^{\Phi} + 4\lambda_3^{\Phi})v^3 + 2(\lambda_1^{\Phi\chi} + \omega^2\lambda_2^{\Phi\chi} + \omega\lambda_2^{\Phi\chi*})vv_{\chi}^2 \nonumber \\
  && + 2(\lambda_1^{\Phi\psi} + \omega\lambda_2^{\Phi\psi} + \omega^{2}\lambda_2^{\Phi\psi*})vv_{\psi}^2
   + (\delta^{\Phi\psi}_{s} - i\delta_a^{\Phi\psi})vv_{\psi} + 2\lambda^{\Phi\vartheta}vv^2_{\vartheta} ~,~\nonumber\\
  0=\frac{\partial V}{\partial\chi_{1}} &=& (2\mu^2_{\chi} + 6\lambda_1^{\Phi\chi}v^2 + 2\lambda_1^{\psi\chi}v^2_{\psi} + 2\omega\lambda_2^{\psi\chi}v^2_{\psi} + 2\omega^{2}\lambda_2^{\psi\chi*}v^2_{\psi})v_{\chi} + 4(\lambda_1^{\chi}+\lambda_2^{\chi})v^3_{\chi} ~,~\nonumber\\
  0=\frac{\partial V}{\partial\chi_{2}} &=& 2\lambda_3^{\Phi\chi}v^2v_{\chi} + \delta^{\chi\vartheta\psi}v_{\psi}v_{\vartheta} ~,~\nonumber\\
  0=\frac{\partial V}{\partial\chi_{3}} &=& 2v_{\chi}(\lambda_3^{\Phi\chi}v^2 + \delta^{\psi\chi}v_{\psi}) ~,~\nonumber\\
  0=\frac{\partial V}{\partial\psi_{1}} &=& 2\delta^{\Phi\psi}_{s}v^2 + 2\lambda_3^{\Phi\psi}v^2v_{\psi} + \delta^{\chi\vartheta\psi}v_{\chi}v_{\vartheta} ~,~\nonumber\\
  0=\frac{\partial V}{\partial\psi_{2}} &=& 2v_{\psi}\{\mu^2_{\psi}+2\lambda_1^{\psi}v^2_{\psi}  + (3\lambda_1^{\Phi\psi}v^2 + \lambda_1^{\psi\chi}v^2_{\chi})
   +  v^2_{\chi}(\omega\lambda_2^{\psi\chi} + \omega^{2}\lambda_2^{\psi\chi*})+\lambda_2^{\psi\vartheta}\vartheta^{2}\}+ 2\delta^{\Phi\psi}_{s}v^2 ~,~\nonumber\\
  0=\frac{\partial V}{\partial\psi_{3}} &=& 2v^2(\delta^{\Phi\psi}_{s} + \lambda_3^{\Phi\psi}v_{\psi}) ~,~\nonumber\\
  0=\frac{\partial V}{\partial\vartheta} &=& 2v_{\vartheta}(\mu^2_{\vartheta} + 2\lambda^{\vartheta}v^2_{\vartheta} + 3\lambda^{\Phi\vartheta}v^2 + \lambda^{\psi\vartheta}v^2_{\psi})~.
 \end{eqnarray}
Note here that for $\lambda^{\eta}>0$ and $\mu^2_{\eta}+\lambda^{\eta\chi}\upsilon^2_{\chi}+\lambda^{\eta\psi}\upsilon^2_{\psi}+ \lambda^{\eta\vartheta}\upsilon^2_{\vartheta}>0$ the new Higgs doublet get a zero VEV, i.e., $\upsilon_{\eta}=0$. For the equations $\partial V/\partial\varphi_{i}$ to be consistent we can force the couplings $\lambda^{\Phi\chi}_{2}, \delta^{\Phi\psi}_{s}$ and $\delta^{\Phi\psi}_{a}$ to vanish. And by forcing $\delta^{\chi\vartheta\psi},\lambda^{\Phi\chi}_{3}$ and $\delta^{\psi\chi}$ to vanish, we obtain $\partial V/\partial\chi_{2,3}=0$, and in this case $\partial V/\partial\psi_{1,3}=0$ are automatically satisfied.
Then we left four independent equations for the four unknown parameters $v$, $v_{\chi}$, $v_{\psi}$, and $v_{\vartheta}$:
\begin{eqnarray}
  0=\frac{\partial V}{\partial\varphi_{i}} &=& 2v\{\mu^2_{\Phi} + 2(3\lambda_1^{\Phi} + 4\lambda_3^{\Phi})v^2 + \lambda_1^{\Phi\chi}v_{\chi}^2 + \lambda_1^{\Phi\psi}v_{\psi}^2 + \lambda^{\Phi\vartheta}v^2_{\vartheta}\} ~,\nonumber\\
  0=\frac{\partial V}{\partial\chi_{1}} &=& 2v_{\chi}\{\mu^2_{\chi} + 3\lambda_1^{\Phi\chi}v^2 + (\lambda_1^{\psi\chi} + \omega\lambda_2^{\psi\chi} + \omega^{2}\lambda_2^{\psi\chi*})v^2_{\psi} + 2(\lambda_1^{\chi}+\lambda_2^{\chi})v^2_{\chi}\} ~,\nonumber\\
  0=\frac{\partial V}{\partial\psi_{2}} &=& 2v_{\psi}\{\mu^2_{\psi} +2\lambda_1^{\psi}v^2_{\psi}+ (3\lambda_1^{\Phi\psi}v^2 + \lambda_1^{\psi\chi}v^2_{\chi}) + v^2_{\chi}(\omega\lambda_2^{\psi\chi} + \omega^{2}\lambda_2^{\psi\chi*})+\lambda_2^{\psi\vartheta}\vartheta^{2}\} ~,\nonumber\\
  0=\frac{\partial V}{\partial\vartheta} &=& 2v_{\vartheta}(\mu^2_{\vartheta} + 2\lambda^{\vartheta}v^2_{\vartheta} + 3\lambda^{\Phi\vartheta}v^2 + \lambda^{\psi\vartheta}v^2_{\psi})~.
 \end{eqnarray}

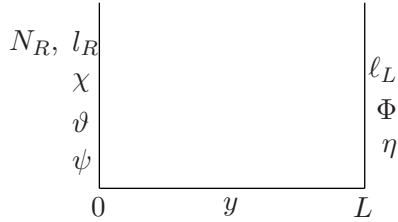
\begin{figure}[ht]
\begin{picture}(300,100)(0,0)
\Line(100,80)(100,10)
\Line(100,10)(200,10)
\Line(200,80)(200,10)
\Text(90,65)[l]{$l_{R}$}
\Text(85,65)[r]{$N_{R},$}
\Text(90,50)[l]{$\chi$}
\Text(90,35)[l]{$\vartheta$}
\Text(90,20)[l]{$\psi$}
\Text(100,3)[c]{0}
\Text(150,3)[c]{$y$}
\Text(200,3)[c]{$L$}
\Text(213,55)[r]{$~~\ell_L$}
\Text(213,40)[r]{$\Phi$}
\Text(213,25)[r]{$\eta$}
\end{picture}
   \caption{ \label{fig:exd} Fifth dimension and locations of scalar and fermion fields.}
 \end{figure}

There is another generic way to prohibit the problematic interactions terms by separating physically between $(\chi,\vartheta,\psi)$ and $(\Phi,\eta)$. Here we solve the vacuum alignment problem by extending the model with a spacial extra dimension $y$, the method was first introduced in Ref.~\cite{Altarelli:2005yp}. We assume the fields live on the 4D brane at $y = 0$ and $y = L$ as shown in Fig.~\ref{fig:exd}. Heavy neutrino masses arise from local operators at $y=0$, on the other hand, charged lepton masses and Yukawa neutrino interactions are realized by non-local effects involving both branes. A detailed explanation of this possibility is beyond the scope of this paper.
Assuming that the trilinear couplings go to zero $(\delta^\psi,\delta^{\psi\chi}$ and $\delta^{\chi\vartheta\psi}\rightarrow0)$, then the potential can be written on the brane at $y=0$  as,
 \begin{eqnarray}
V(\chi) &=& \mu^2_\chi B_{2} + \lambda^\chi_1 Q^{\chi}_{1} + \lambda^\chi_2 Q^{\chi}_{2}+ \lambda^\chi_3 Q^{\chi}_{3} \\
V(\psi) &=& \mu^2_\psi B_{3} +\lambda^\psi_1 Q^{\psi}_{1} + \lambda^\psi_2 Q^{\psi}_{2}+ \lambda^\psi_3 Q^{\psi}_{3} \\
V(\vartheta) &=& \mu^2_\vartheta B_{4} + \lambda^\vartheta Q^{\vartheta}_{1}\\
V(\psi, \chi) &=& \lambda^{\psi\chi}_{1}Q^{\psi\chi}_{1}+\lambda^{\psi\chi}_{2}Q^{\psi\chi}_{2}+\lambda^{\psi\chi\ast}_{2}Q^{\psi\chi}_{3} +\lambda^{\psi\chi}_{3}Q^{\psi\chi}_{4}\\
V(\psi, \vartheta) &=& \lambda^{\psi\vartheta}Q^{\psi\vartheta}_{1}
 \end{eqnarray}
 and on the brane $y=L$,
 \begin{eqnarray}
 V(\Phi) &=& \mu^2_\Phi B_{1} +\lambda^\Phi_1 Q^{\Phi}_{1} + \lambda^\Phi_2 Q^{\Phi}_{2}+\lambda^\Phi_3 Q^{\Phi}_{3} + \lambda^\Phi_4 Q^{\Phi}_{4}+ i \lambda^\Phi_5 Q^{\Phi}_{5}\\
V(\eta) &=& \mu^2_\eta B_{5} + \lambda^\eta Q^{\eta}_{1}
\\
V(\Phi, \eta) &=& \lambda^\eta Q^{\eta}_{1}+ \lambda^{\Phi\eta}_1 Q^{\Phi\eta}_{1} + \lambda^{\Phi\eta}_2 Q^{\Phi\eta}_{2} + \lambda^{\Phi\eta}_3 Q^{\Phi\eta}_{3} +\lambda^{\Phi\eta*}_3 Q^{\Phi\eta}_{4}.
 \end{eqnarray}
 The minimal conditions of potential $V_{y=0}$ are
 \begin{eqnarray}
 \frac{\partial V_{y=0}}{\partial \chi_1} &=& 2v_{\chi}\{\mu^2_{\chi} + (\lambda_1^{\psi\chi}+ \omega\lambda_2^{\psi\chi}+ \omega^{2}\lambda_2^{\psi\chi*})v^2_{\psi}+ 2(\lambda_1^{\chi}+\lambda_2^{\chi})v^2_{\chi}\} = 0, \\
 \frac{\partial V_{y=0}}{\partial \psi_2} &=& 2v_{\psi}\{\mu^2_{\psi} +2\lambda_1^{\psi}v^2_{\psi}+ \lambda_1^{\psi\chi}v^2_{\chi} + (\omega\lambda_2^{\psi\chi} + \omega^{2}\lambda_2^{\psi\chi*})v^2_{\chi}+\lambda^{\psi\vartheta}v^2_{\vartheta}\} = 0, \\
 \frac{\partial V_{y=0}}{\partial \vartheta} &=& 2v_{\vartheta}\{\mu^2_{\vartheta} + 2\lambda^{\vartheta}v^2_{\vartheta} + \lambda^{\psi\vartheta}v^2_{\psi}\} = 0,
 \end{eqnarray}
  and $\frac{\partial V_{y=0}}{\partial \chi_{2,3}} = \frac{\partial V_{y=0}}{\partial \psi_{1,3}} = 0$ are automatically satisfied. While the minimal conditions on the brane $y = L$ are
 \begin{eqnarray}
 \frac{\partial V_{y=L}}{\partial \varphi_i} = 2v\{\mu^2_{\Phi}+ 2(3\lambda_1^{\Phi} + 4\lambda_3^{\Phi})v^2\} = 0 \quad (i = 1,2,3).
 \end{eqnarray}
 We left four independent equations for the four unknown $v$, $v_{\chi}$, $v_{\psi}$, and $v_{\vartheta}$ in the limit of the trilinear couplings going to be zero. Thus the configurations needed in our scenario can be realized with {\it ad hoc} constraints.


\end{document}